
\baselineskip=16pt
\pageno=0
\rightline{ILL-(TH)-92-\#19}
\rightline{CERN-TH.6557/92}
\vskip0.5 truecm
\centerline{\bf FOUR - FERMI THEORIES IN FEWER THAN FOUR DIMENSIONS}
\vskip 0.5 truecm
\centerline{{\bf Simon Hands}}
\centerline{\it Department of Physics and Astronomy,
University of Glasgow,
Glasgow G12 8QQ, U.K.}
\centerline{\it and}
\centerline{\it Theory Division,
CERN,
CH-1211 Geneva 23, Switzerland}
\vskip 0.5 truecm
\centerline{{\bf Aleksandar Koci\'c}}
\centerline{\it Theory Division,
CERN,
CH-1211 Geneva 23, Switzerland}
\vskip 0.5 truecm
\centerline{{\bf John B. Kogut}}
\centerline{\it Department of Physics,
University of Illinois at Urbana-Champaign,}
\centerline{\it 1110 West Green Street, Urbana, IL 61801-3080, U.S.A.}
\vskip 1.0 truecm
\centerline{\bf Abstract}\footnote{$\,^{}$}{CERN-TH.6557/92}

{\narrower
\noindent
Four-fermi models in dimensionality $2<d<4$ exhibit an ultra-violet stable
renormalization group fixed point at a strong value of the coupling constant
where chiral symmetry is spontaneously broken. The resulting field theory
describes relativistic fermions interacting non-trivially via exchange of
scalar bound states. We calculate the $O(1/N_f)$ corrections to this picture,
where $N_f$ is the number of fermion species, for a variety of models and
confirm their renormalizability to this order. A connection between
renormalizability and the hyperscaling relations between the theory's critical
exponents is made explicit. We present results of extensive numerical
simulations of the simplest model for $d=3$, performed using the hybrid Monte
Carlo algorithm on lattice sizes ranging from $8^3$ to $24^3$.
For $N_f=12$ species of massless fermions we confirm the existence of a
second order phase transition where chiral symmetry is spontaneously broken.
Using both direct measurement and finite size
scaling arguments we estimate the critical exponents $\beta$, $\gamma$, $\nu$
and $\delta$. We also investigate symmetry restoration at non-zero temperature,
and the scalar two-point correlation function in the vicinity of the bulk
transition. All our results are in excellent agreement with analytic
predictions, and support the contention that the $1/N_f$ expansion is accurate
for this class of models.
\smallskip}
\footnote{$\,^{}$}{August 1992}
\vfill\eject
%
%
%
\baselineskip=20pt
\parskip= 5pt plus 1pt
\parindent=15pt
\noindent{\bf1. Introduction}

The existence of a continuum limit of four-fermi theories is a long standing
problem that goes back to the work of Nambu and Jona-Lasinio [1].
The original motivation for studying these
models was the fact that they give a qualitatively good description of
dynamical chiral
symmetry breaking in strong interaction physics. The only degrees of freedom
involved are fermions which interact via a short range interaction. As such
four-fermi theories do not have a renormalizable perturbation expansion in
powers of the coupling constant.
However, at strong couplings, scalar bound states appear in the
spectrum as a consequence of chiral symmetry breaking and the
fermions interact by exchanging these composite scalars. The appearance of the
scalars and pseudoscalars is tied to the Nambu-Goldstone realisation of chiral
symmetry. By symmetry arguments, it
can be seen that the emerging low-energy
theory is similar to the linear $\sigma$-model, with scalars coupled to
fermions via a Yukawa interaction,
which is known to be
perturbatively renormalizable. In four dimensions,
the equivalence between the
two models was noted very early on [2] and has been exploited by many
authors [3,4]. Meanwhile, it was discovered by Gross and Neveu that in
two dimensions the model is asymptotically free and has a nontrivial
continuum limit near the origin [5]. Wilson has argued
that four-fermi theories
are nontrivial for all $d<4$ [6]. His arguments are based on the leading order
results of the $1/N_f$ expansion, where $N_f$ is the number of fermion species
in the model.
Many other studies have been done using this expansion
technique [7]; a series
of articles by Guralnik et al. [3] indicated that below four dimensions,
an equivalence
between four-fermi and sigma models could be established.

For definiteness, let us introduce the model we shall be dealing with in the
bulk of this article. It is often called the Gross-Neveu model in the
literature, and is the simplest relativistic theory of interacting fermions.
The continuum space-time Lagrangian (we work in Euclidian space throughout)
reads,
$${\cal L}=\sum_{j=1}^{N_f}\left[
\bar\psi_j
(\partial{\!\!\! /}\,+m)\psi_j-{g^2\over{2N_f}}(\bar\psi_j\psi_j)^2
\right],\eqno(1.1)$$
where the index $j$ labels the $N_f$ species, each of which is described by a
four component spinor.  For purposes of analysis as well as computation, it
is useful to introduce an auxiliary field $\sigma$ so that ${\cal L}$ becomes
quadratic in $\psi_j$:
$${\cal L}=\sum_{j=1}^{N_f}\left[
\bar\psi_j
(\partial{\!\!\! /}\,+m)\psi_j+\sigma\bar\psi_j\psi_j+{N_f\over{2g^2}}
\sigma^2\right].\eqno(1.2)$$
Gaussian functional integration over the $\sigma$ field in (1.2) restores the
original Lagrangian (1.1) -- however, it also serves as an
interpolating field for the scalar bound state governing the fermion
interactions described above. In the chiral limit
fermion bare mass $m\to0$
the vacuum expectation value of $\sigma$ becomes a dynamical
fermion mass and is a convenient order parameter for any chiral symmetry
transition in the theory.  Note that in this limit
Eq. (1.2) has a discrete ${\rm Z}_2$
chiral symmetry: $\sigma\mapsto-\sigma$; $\psi_j\mapsto\gamma_5\psi_j$;
and $\bar\psi_j\mapsto-\bar\psi_j\gamma_5$. It is this symmetry which
is spontaneously broken at strong coupling. The universal critical indices of
this critical point have been calculated to $O(1/N_f)$: for $d=3$, they read
$$
\nu=1+{8\over{3N_f\pi^2}};\;\;\delta=2+{8\over{N_f\pi^2}};\;\;
\beta=1+O({1\over N_f^2});\;\;\gamma=1+{8\over{N_f\pi^2}};\;\;
\eta=1-{16\over{3N_f\pi^2}},\eqno(1.3)$$
in the conventional notation of classical statistical mechanics [8].  These
indices are particularly interesting because they are far from free-field
Gaussian model indices.  Apparently, the critical field theory describes a
strongly interacting fermionic system, complete with composite states, an
interesting S-matrix, etc.  The critical indices Eq. (1.3) lie at the heart
of this article.  The deviation of the critical indices from the Gaussian
model indices are related to the anomalous dimensions of the various
composite fields in the model, as we shall now discuss.

Below four dimensions the Yukawa theory is
superrenormalizable with an ultraviolet fixed point at the
origin
and an infra red fixed point at strong couplings, where we
recover scale invariance with scaling governed by non-vanishing
anomalous dimensions. At this point the scalar degrees of freedom become
composite and the low energy theory is the same as that of the four-fermi
theory
in the UV region. The compositeness condition is manifested in the vanishing
of the scalar kinetic term in the continuum limit. Whereas
at weak coupling $(\partial_\mu \sigma)^2$
is marginal irrespective of $d$, in the vicinity of the infrared fixed point
the field
$\sigma$ scales with an anomalous dimension which, because of unitarity, must
be positive. This makes the kinetic term irrelevant at the IR fixed point.
Since the IR limit of
the Yukawa model now has the same relevant operator content
as the UV limit of the
four-fermi model, the latter must be renormalizable.
Recently, several groups have studied the three dimensional
model and demonstrated that Wilson's ideas survive beyond the leading
order in $1/N_f$. In particular,
a series of papers by Rosenstein {\it et al\/} [9] have
pointed out that the $1/N_f$ expansion around the fixed point
in $d=3$ is renormalizable. This result was extended to general $d\in(2,4)$
in [10]. Explicit $O(1/N_f)$ calculations appear in references [8,11].
Ref.[12] contains a rigorous proof of renormalizability of the
three-dimensional theory.

We should mention the physical meaning of the compositeness condition
$Z=0$, where $Z$ is the wavefunction renormalization factor associated with
the scalar kinetic term,
and its relationship with the fixed-point condition. Consider the general
case of an interacting theory where a bound state $|B>$ with binding energy
$E_B =-B<0$ appears. The wavefunction renormalization constant is defined by
$$
Z=\sum_b |<b|B>|^2
\eqno(1.4)
$$
where $|b>$ stands for the bare (elementary particle) states. Using
standard techniques, it can be shown that $Z$ satisfies the following equation
[13]
$$
1-Z=\int_0^\infty d\epsilon { {G^2(\epsilon)}\over{(\epsilon+B)^2} }
\eqno(1.5)
$$
with $G^2(\epsilon)$ being the total decay rate of the state $|B>$
proportional to some effective coupling constant. From Eq. (1.4),
we see that $Z=0$ means that the bound state has no projection on the space
of bare states. This is the compositeness condition.
{}From (1.5) it is transparent that this condition
puts an upper limit on the effective coupling $G(\epsilon)$ thus yielding
fixed-point condition.

The renormalizability of a given theory can be argued in several ways. One is
the old-fashioned way of constructing a finite theory by adding counterterms
and ensuring that all divergences cancel order by order in some kind of
expansion. Another is to start with a discretized, and therefore UV finite,
theory and look for the region in the bare parameter space where a macroscopic
length scale appears. The two methods are equivalent and we will use them
both in this paper.
The general idea of renormalizability is that the cutoff dependence can be
absorbed
into a finite set of bare parameters in such a way that
the low energy physics which emerges
is insensitive to the cutoff. Once this is done, it is possible to
find lines of constant physics -- renormalization group (RG) trajectories --
in the bare parameter space. These lines are uniquely defined no matter
which observable is taken.
This way, the low-energy quantities depend on each
other and not on the cutoff -- we will see that this is possible if certain
consistency conditions known as hyperscaling relations, governing the scaling
of
different physical quantities, are obeyed.

The hyperscaling hypothesis [14] claims that the only relevant scale in
the critical region is the macroscopic correlation length
$\xi$. If this hypothesis holds, it is
possible to do dimensional analysis using this correlation length as a scale.
As a consequence, all dimensionless  quantities should be independent of $\xi$.
In the context of a simple model of ferromagnetism, the
hypothesis can be stated as
$$
F_{sing}=t^{2-\alpha} F(h/t^\Delta)\propto\xi^{-d}
\eqno(1.6)
$$
where $F_{sing}$ is the singular piece of the free energy density and
$\alpha, \Delta=\beta\delta$
are specific heat and gap exponents respectively. The external
symmetry breaking field is $h$ and $t=T-T_c$ is the deviation from the critical
temperature. All thermodynamic quantities can be obtained by taking
the derivatives of the free energy. In particular, the
order parameter (magnetization), defined as
$<\phi>=\partial F_{sing}/\partial h$, satisfies the equation of state (EOS)
$$
<\phi>=t^{2-\alpha -\Delta}F'(h/t^\Delta) \equiv t^\beta F'(h/t^\Delta)
\eqno(1.7)
$$
with the magnetic exponent defined as $\beta =2-\alpha-\Delta$.
Similarly, the susceptibility exponent, $\gamma$, is defined via
$$
\chi\equiv
{{\partial <\phi>}\over{\partial h}}
= t^{\beta-\Delta} F''(h/t^\Delta) \equiv
t^{-\gamma} F''(h/t^\Delta)
\eqno(1.8)
$$
i.e. $\gamma=\Delta-\beta$.
The behavior of the correlation length in the scaling region is given by
$$
\xi=t^{-\nu} g(h/t^\Delta)
\eqno(1.9)
$$
Using the hyperscaling hypothesis (1.6) we relate $\alpha$ to $\nu$:
$d\nu=2-\alpha$.
The exponents $\beta$ and $\nu$ can be related to the scaling dimension of
the field $\phi$. If a field $\phi$, which develops nonvanishing
vacuum expectation value $<\phi>\not=0$, has scaling dimension
$d_\phi\equiv{1\over2}(d-2+\eta)$, then
hyperscaling implies $<\phi>\sim \xi^{-d_\phi}$. The vanishing
of the order parameter is given by $<\phi>\sim t^\beta$ from where it follows
that $\beta/\nu =d_\phi$. The other scaling relations are derived similarly.
For a given channel, the corresponding correlation length (= inverse mass)
may be defined by
$$2d\xi_\phi^2={{\int_x |x|^2<\phi(x)\phi(0)>}\over
{\int_x<\phi(x)\phi(0)>}}
\eqno(1.10)
$$

Since hyperscaling is an important statement, we should explain its meaning and
outline possible implications.
It is generally believed that the violation of hyperscaling leads to
triviality. This is due to inequalitites between
certain combinations of critical
indices [15]. In the simplest cases like scalar field theories and spin
systems, the quantity that measures the violation of hyperscaling is
the {\it dimensionless}
renormalized coupling. It is defined through the nonlinear susceptibility
$\chi^{(nl)}$:
$$
g_R=-{ {\chi^{(nl)}}\over{\chi^2 \xi^{d}} }, \,\,\,\,\,\,
\chi^{(nl)}={ {\partial^3 <\phi>}\over{\partial h^3} }
\eqno(1.11)
$$
The renormalized coupling is essentially the properly normalized connected
four-point function. In terms of the correlation functions, the
nonlinear susceptibility is the zero-momentum projection of the four-point
function
$$
\chi^{(nl)} = \int_{xyz}<\phi(x)\phi(y)\phi(z)\phi(0)>_{conn}
\eqno(1.12)
$$
The normalization factors can be understood by recalling that susceptibility
and mass are related to the wave function normalization constant $Z$ through
$Z=\chi/\xi^2$. In the numerator we get one power
of $Z^{1/2}$ for each field, which is thus compensated by $\chi^2$: the
factor of $\xi^d$ accounts for the fact that there is one extra spatial
integration in the numerator of (1.11). Both $\chi^{(nl)}$ and $\chi$ can be
obtained by differentiating the free energy (1.6): $\chi=t^{2-\alpha-2\Delta}
F^{\prime\prime}(y)$ and $\chi^{(nl)}=t^{2-\alpha-4\Delta}F^{iv}(y)$. Together
with the EOS for the correlation length (1.9), they give the expression for the
renormalized coupling
$$g_R=t^{d\nu-(2-\alpha)}H(h/t^\Delta),\eqno(1.13a)$$
or, using the definition of the exponents $\beta$ and $\gamma$, Eqs. (1.7,8),
the alternative form
$$
g_R = t^{-2\Delta+\gamma+d\nu} H(h/t^\Delta)
\eqno(1.13b)
$$
In specific (ferromagnetic) models an estimate of the exponent can be made.
Since multi-spin correlations cannot extend over a larger range than pair
correlations, the following inequality holds [15]:
$$
2\Delta\leq \gamma +d\nu
\eqno(1.14)
$$
We can trade $t$ with the correlation length in Eq.(1.9) and rewrite Eq.(1.13)
in terms of $\xi$ as
$$
g_R = \xi^{(2\Delta-\gamma-d\nu)/\nu} {\tilde H}(h/t^\Delta)
\eqno(1.15)
$$
Since $g_R$ is dimensionless, hyperscaling implies that it must be
independent of $\xi$ and the exponent must vanish. In this case the
renormalized
coupling is a function of only one bare variable. Strict inequality in
Eq.(1.14)
implies triviality. An equivalent way of stating the
above inequality is $d\nu\geq 2-\alpha$.
It implies that the singular part
of the free energy vanishes no faster than the prediction of hyperscaling
i.e. $F_{sing}\geq \xi^{-d}$.

In a similar fashion the scaling of mass ratios can be derived [16].
If hyperscaling is satisfied, for
any particular pair of masses that obey Eq.(1.9) e.g. $M_\pi, M_\sigma$,
we have
$$
R(t,h)={ {M_\pi^2}\over{M_\sigma^2} }=
G(h/t^\Delta)
\eqno(1.16)
$$
where $G(y)$ is a universal function. Comparing it with the expression for
the renormalized coupling, we see that both observables depend on the same
variable.  One of the relations, $R=G(y)$ or $g_R=H(y)$
can be inverted to solve for the bare variable
e.g. $y=H^{-1}(g_R)$.
This defines an RG trajectory for each value of the renormalized coupling:
$h=H^{-1}(g_R)t^\Delta$.
This is then used to obtain the relation between the two observables
$R=R(g_R)$. The same manipulation can be done with two mass ratios.
We note
that the important point in the inversion is that both observables depend
on just one bare variable, so that the inverse relation can always be found,
at least in some regions of parameter space. This would be difficult to
achieve otherwise, and is a consequence of hyperscaling.

To summarize, the requirement of having a macroscopic correlation
length as the only relevant scale leads to relations between the critical
exponents. These relations are
$$
2-\alpha =d\nu, \,\,\,\,\,\,
2\Delta=\gamma+d\nu, \,\,\,\,\,\,
\beta={\nu\over 2}(d-2+\eta), \,\,\,\,\,\,
\gamma=\nu(2-\eta)
\eqno(1.17)
$$
All the other relations are obtained from Eqs.(1.17) using the definitions of
the critical exponents e.g. $2\beta\delta-\gamma=d\nu$, $2\beta+\gamma=d\nu$.

Next we  review briefly how these ideas are realized in four-fermi theories,
at least to leading order in $1/N_f$. We shall see in the next section, where
detailed calculations are presented, that subleading corrections do not change
the physics qualitatively.
The critical exponents for the model (1.1) have been calculated in Ref. [8].
The physical fermion mass
$M$ satisfies a gap equation,
$$
M=m-g^2<\bar\psi\psi>,
\eqno(1.18)
$$
In the chiral limit $M$ is proportional to the order parameter
$<\bar\psi\psi>$.
In the large $N_f$ limit, only the simple fermion tadpole contributes to
$<\bar\psi\psi>$.
The exponents $\beta,\delta$ can be obtained from the gap equation.
It is convenient to use the definition of the critical coupling
$1=4g^2_c\int_q 1/q^2$, which is cutoff dependent,
to rewrite the gap equation in the form
$$
tM+m=4g^2\int_q { {M^3}\over{q^2(q^2+M^2)} }
\eqno(1.19)
$$
where $t=(g^2-g_c^2)/g_c^2$. Below four dimensions, the integral is IR
divergent in the absence of the fermion mass $M$. It can be evaluated
exactly and it is of the order of  $M^{d-1}$. Thus, in the
limit when either bare mass $m$ or $t$ vanish, we have
$$\eqalign{
M&\sim t^{1/(d-2)}  \,\,\,\,\,\,\,\, (m=0),\cr
M&\sim m^{1/(d-1)}  \,\,\,\,\,\,\,\, (t=0).\cr}
\eqno(1.20)
$$
Since $<\bar\psi\psi>\sim M$, the exponents are:
$\beta=1/(d-2)$, $\delta=d-1$.

The exponents $\nu,\gamma,\eta$ are obtained from the
scalar propagator $D_\sigma$. To leading order in $1/N_f$ it is given by
$$
D_\sigma^{-1}(k^2)=1+g^2\Pi(k^2),
\eqno(1.21)
$$
where $\Pi(k^2) = - {\rm tr} \int_q S(q+k) S(q)$, and $S$ is the fermi
propagator $(ip{\!\!\! /}\,+M)^{-1}$.
We associate the scalar mass with the inverse correlation length
$1/M_\sigma^2=\xi^2$. The susceptibility and
wavefunction renormalization constant for the
scalar field are: $\chi^{-1}= 1+g^2\Pi(0),\,\,\,\, Z^{-1}=g^2\Pi'(0)$, which
using (1.10) leads to
$$
\xi^2=\Biggl({1\over
{D_\sigma^{-1}(k^2)}}
{ {dD_\sigma^{-1}(k^2)}\over
{dk^2}}\Biggr)_{k^2=0}
\eqno(1.22)
$$
The scalar mass and renormalized coupling defined in this way are thus
$$
M_\sigma^2=Z\chi^{-1}={ {1+g^2\Pi(0)}\over{g^2\Pi'(0)}},
\,\,\,\,\,\,\,\,\,\,
\tilde g^2_R= Z g^2 ={1\over {\Pi'(0)} },
\eqno(1.23)
$$
where the renormalized Yukawa coupling $\tilde g_R$ is {\it dimensionful\/}.
It has dimension $(mass)^{(4-d)/2}$ and,
since it is a low-energy quantity, it depends on the physical
mass $M_\sigma$. The {\it dimensionless\/} coupling is thus
$$
g^2_R={{\tilde g^2_R}\over{M_\sigma^{4-d}}}.
\eqno(1.24)
$$

The central question in the study of four-fermi
theories is whether the bound states remain composite or become pointlike
in the continuum limit.
An important consequence of compositeness is the non-triviality of the theory;
ie. the dimensionless interaction strength
remains non-vanishing in the continuum limit.
The physical reason in the case of the linear $\sigma$-model
is easy to understand. The generic coupling between the fermions and scalars is
given by the Goldberger-Treiman relation: $g_{\pi f\bar f}=M/f_\pi$,
where $M$ is the fermion dynamical mass and
$f_\pi$ is the pion decay constant related to the pion radius
by $r_\pi\sim 1/f_\pi$. This way, the Yukawa coupling becomes
$g_{\pi f\bar f}\sim M r_\pi$ and vanishes in the limit of pointlike pions.
In other words, since the scalar mass is not protected from receiving
large corrections (of the order of the cutoff scale) via radiative
corrections, the effective interactions will always be zero-range in
the continuum limit.
Such a theory does not have a physical scale and must be trivial
above two dimensions. The only way to have a nontrivial limit is if
the exchanged particles are composite. We will see later that this requirement
is achieved when the bare parameters are tuned so as to make all
power-law divergences disappear.

We restrict the discussion to $2<d<4$ in what follows to avoid
violations of scaling although, regarding the flow of RG trajectories,
the conclusions will be the same
for $d=2$ (but not for $d=4$). Using the gap equation, we get
(the quantity $\Pi(k^2)$ will be evaluated in detail in the next section)
$$
\chi^{-1}={m\over M}+{12\over(d-1)}M^2 Z^{-1},
\,\,\,\,\,\,\,\,
M_\sigma^2 = {m\over M}Z+{12\over(d-1)}M^2,
\,\,\,\,\,\,\,\,
Z^{-1}= {{bg^2}\over M^{4-d}},
\eqno(1.25)
$$
where $b=2\Gamma(2-d/2)(d-1)/3(4\pi)^{d/2}$.
We measure all the masses in units of the momentum cut-off.
Note that the wavefunction renormalization constant $Z$ vanishes in the
continuum limit $M\to0$: this is precisely the compositeness condition [13].
In the chiral limit $m\to0$ the
susceptibility is $\chi^{-1}\propto M^{d-2}$, which
implies, using Eq.(1.20), that $\chi^{-1}\propto t$, i.e. $\gamma=1$.
Similarly,
the $\sigma$ mass in Eq.(1.25) scales as $M_\sigma\propto M\propto t^{d-2}$
giving $\nu=1/(d-2)$. The exponent $\eta$ is extracted from the power law decay
of the scalar correlation function at the critical point. Simple power
counting in Eq.(1.21) gives
$$
\displaystyle\lim_{k^2\to\infty}
D_\sigma(k^2)\propto {1\over{ k^{d-2}}}\propto{1\over{k^{2-\eta}}},
\eqno(1.26)
$$
which gives $\eta=4-d$.

To obtain the RG trajectories and relate the low energy observables,
we use Eq.(1.25) to calculate the mass ratio $R=M_\sigma^2/M^2$ and
the renormalized coupling:
$$
R={1\over{b g^2}} {m\over M^{d-1}}+{12\over(d-1)},
\,\,\,\,\,\,\,\,
g_R^2={1\over{b R^{2-d/2}}}.
\eqno(1.27)
$$
It is clear from the second relation that the lines of constant mass ratio
are also lines of constant renormalized coupling.
Regarding the ratio between the scalar and fermion masses
we observe two things. In the limit $d\to4$, since $1/b\propto4-d$,
$R\to 4$, which is the expected result if the scalar is a weakly bound
fermion -- anti-fermion state. For arbitrary $d$ in the chiral limit, we
obtain the result $R=R_c=12/(d-1)$. At first sight this seems to contradict the
interpretation of $\sigma$ as a weakly bound state -- however, this result
is due to our using the definition (1.22) for the scalar mass, together with
the fact that the numerical coefficient of $Z$ is far from unity.
As we shall confirm in the next section, the scalar propagator has a pole at
$k^2=-4M^2$ in the chiral limit for all $d$. Away from the chiral limit,
however, $D_\sigma$ no longer has a pole, which means that the
definition (1.22) is more useful from our perspective.
Secondly, consider the gap equation;
$$
c g^2 M^{d-2}=
t+{m\over M},
\eqno(1.28)
$$
with $c=8\Gamma(2-d/2)/(d-2)(4\pi)^{d/2}$.
To leading order in $1/N_f$ this can be
regarded as the EOS.
{}From the expression for $R$ and the gap equation we eliminate $M$ to obtain
$$
m=t^{\Delta}
{1\over{g^{2\beta}}}
{
{b(R-R_c)}\over
{\left(c-b(R-R_c)\right)^\Delta}
},
\eqno(1.29)
$$
where $\Delta=\beta\delta=(d-1)/(d-2)$. This is the expected
scaling of the RG trajectories $m\sim t^\Delta$. The lines of constant $R$ fall
into two subfamilies. For $t>0, R-R_c<c/b$ and $t<0, R-R_c>c/b$. In the chiral
limit $R=R_c$.
As the binding increases, the mass of the composite, measured
in units of the constituent mass, decreases.

We comment about the theory above four dimensions where
hyperscaling is violated and where, as a consequence, the continuum limit
is trivial. It is easy to see that, for $d\geq 4$ the gap equation (1.19)
is regular in the $M\to 0$ limit. As a consequence, the corresponding
exponents $\beta,\delta$ are the same as in four dimensions; $\beta=1/2$,
$\delta=3$. Again, the reason for this is the IR behavior of the theory.
The gap equation in this case is
$$Cg^2\Lambda^{d-4}M^3=tM+m, \eqno(1.30)$$
where we display the explicit cutoff dependence, since the physical mass will
turn out not to be the only scale. It is easy to verify that the expression
(1.29) for the mass ratio will remain unchanged. However, the renormalized
coupling will change;
$$g_R^2=ZM^{d-4}g^2={1\over B}\left({M\over\Lambda}\right)^{d-4}.\eqno(1.31)$$
The wavefunction renormalization constant is obtained
as the $k^2$ coefficient in the scalar propagator. To leading order it is
$$
Z^{-1} \sim \int_q { 1\over{(q^2+M^2)^2} },
\eqno(1.32)
$$
which is clearly finite in the $M\to 0$ limit for $d>4$.
The relations for the remaining critical exponents follow from Eqs.(1.25):
$\gamma=1$, $\nu=1/2$, $\eta=0$. For $d>4$ hyperscaling is clearly violated:
indeed with mean field values of the exponents Baker's inequality
(1.14) becomes $d\geq4$.
Two things are clear: firstly the trajectories of constant $R$ and $g_R$ do
not coincide; and secondly in the limit $M/\Lambda\to0$, the renormalized
coupling vanishes. Because of the
violation of hyperscaling the theory is trivial.

In four dimensions, hyperscaling violations are logarithmic and are not
reflected through the violation of the relations between the critical
exponents.
Explicit calculation to leading order gives
$$
g_R^2={ {8\pi^2}\over{\ln\left({\displaystyle{\Lambda^2\over m^2}}\right)} }
\eqno(1.33)
$$
So, the theory is trivial, but the critical exponents satisfy hyperscaling
relations. The reason for this is that the scalars are pointlike: $Z^{-1}$
is logarithmically divergent in the
UV cutoff.

In section 2 of this article we present a
calculation of $1/N_f$ corrections to the model at length, for arbitrary
dimensionality $2<d<4$. We will explicitly demonstrate the renormalizability
of the $1/N_f$ expansion to next-to-leading order in this regime, and also
calculate the critical indices and show that they continue to satisfy
the hyperscaling relations. We will argue that the
physical assumptions underlying renormalizability and hyperscaling are
equivalent, and demonstrate this using the chiral symmetry of (1.1,2). For
completeness we also generalize our results to models where the spontaneously
broken symmetry is ${\rm U}(1)\otimes{\rm U}(1)$ or
${\rm SU}(2)\otimes{\rm SU}(2)$. Some of these results were presented in
briefer form in Ref. [8].

In the rest of the article we shall fix $d=3$ and
study the model by computer
simulation methods.  There are several motivations for doing this.  First,
we can investigate the validity of the $1/N_f$ expansion by a truly
non-perturbative numerical scheme.  We shall see that standard methods of
extracting critical behavior from computer simulations yield results in
good agreement with the $1/N_f$ expansion when $N_f$ is chosen large --
$N_f$ will be
set to 12 in our numerical work.  Second, we can study the model for small
$N_f$, where it may have some relevance to the strongly correlated electron
dynamics underlying high $T_c$ superconductivity, and see if there are
dramatic changes as $N_f$ decreases.  Although we have some results at small
$N_f$, this article will concentrate on the first objective and extract a
number of results from $N_f=12$ simulations which provide support for the
usefulness and reliability of the $1/N_f$ expansion. Our main results are
a numerical verification of the critical index predictions of Eq.(1.3) to
leading order.

After describing the lattice
formulation of the model and the hybrid Monte Carlo algorithm in Sec. 3, we
turn to numerical results. Most of our simulations are performed directly in
the chiral limit $m=0$. In Sec. 4 we consider the model for $N_f=12$ on
symmetric lattices ranging in size from $8^3$ to $20^3$.  Measurements of the
vacuum expectation value
$\Sigma_0\equiv<\sigma>$ clearly show a chiral transition in the
lattice model at an inverse coupling $1/g^2$ near unity.  Measurement of
$\Sigma_0$ and its susceptibility
$\chi$ as a function of the coupling $1/g^2$ allow us to measure the
critical indices $\beta$ and $\gamma$ directly. The leading order terms in
Eq. (1.3), $\beta=\gamma=1$, will be confirmed with good precision.
The size dependence of
the critical point will also provide a rough estimate of $\nu$, using finite
size scaling arguments, but more quantitative results will be presented in
later sections of this article.  In particular, the model will be
considered at nonzero temperature in Sec. 5 by simulating it on asymmetric
lattices, $N_\tau\times N^2$
with $N=36$ lattice spacings and $N_\tau$ ranging from 2 through 12.  We
shall find a second order phase transition at a critical temperature in
good agreement with large $1/N_f$ calculations.  A measurement of the critical
index $\nu$ will follow.  In Sec. 6 we introduce a small
bare fermion mass $m$ into the theory's action and measure the
critical index $\delta$ at the zero temperature critical point as well as at
the
finite temperature transition.  The first set of measurements, done on $8^3$,
$16^3$ and $24^3$ lattices, are nicely consistent with the leading order term
in Eq.(1.3), $\delta=2$.  The second set of measurements uses a
$12\times36^2$ lattice and
gives $\delta^T=3.0$, the expected Gaussian model index for the
temperature-driven phase transition [17]. In Sec. 7 we turn to measurements of
the two-point scalar correlation function, and our attempts to extract the
scalar mass using the formalism developed in Sec. 2. As we shall see, this
requires some novel fitting techniques, and the results presented here are
probably only a first step towards a complete understanding. Our results are
summarized in Sec. 8.
\vskip 1.5 truecm
\noindent{\bf2. Calculations of $O(1/N_f)$ Corrections}

In this section we will calculate the lowest non-trivial corrections, ie. those
of $O(1/N_f)$, to the model in the vicinity of the strongly-coupled fixed point
at $g=g_c$. We will work in dimensionality $d$ where $2<d<4$, and
have chosen to follow the treatment of Gat et al [11], and define the
Lagrangian involving the auxiliary scalar field in the chiral limit
as follows:
$${\cal L}=Z_\psi\bar\psi_i\partial{\!\!\! /}\,\psi_i
+{g\over{\sqrt{N_f}}}Z_\psi Z_\sigma^{1\over2}\sigma\bar\psi_i\psi_i
+{1\over2}Z_\sigma \sigma^2,\eqno(2.1)$$
where a sum on $i=1,\ldots,N_f$ is understood. Notice that in this section of
the paper the $\sigma$ field is normalized differently to that of Eq.(1.2);
since $\sigma$ is auxiliary this has no physical consequences.
The constants $Z_\psi$,
$Z_\sigma$, and $g$ are cutoff-dependent, and must be adjusted at each order of
the $1/N_f$ expansion to keep Green functions finite, the relevant ones being
the fermi
propagator $S_F$, the scalar-fermion vertex $\Gamma_{\sigma\bar\psi\psi}$, and
the scalar propagator $D_\sigma$. It should be noted that the constants
$Z_\psi$, $Z_\sigma$ are chosen dimensionless
for convenience and are not directly related to the constant
$Z$ of the previous section, which has dimension $(mass)^2$. For additional
simplicity we work for the most part in the chiral limit $m=0$.
In the broken phase $t>0$ we will choose renormalization
conditions so as to keep the physical fermion mass, that is, the pole of the
renormalized fermion propagator, a finite constant independent of $1/N_f$.

Our Feynman rules are thus:
$$\eqalign{S_F^{(0)-1}&=Z_\psi(ip{\!\!\! /}\,+\Sigma_0);\cr
\Gamma^{(0)}_{\sigma\bar\psi\psi}
&=-{g\over\sqrt{N_f}}Z_\psi Z_\sigma^{1\over2};\cr}
\eqno(2.2)$$
where the dynamical
fermion mass $\Sigma_0$ appearing in the propagator is
{\it defined\/} by the expectation value
of the scalar field (this is equivalent to the definition of $\Sigma_0$ used
elsewhere in the paper):
$$\Sigma_0={{gZ_\sigma^{1\over2}}\over\sqrt{N_f}}<\sigma>.\eqno(2.3)$$
We determine  $\Sigma_0$ from the equation of motion
$\delta{\cal L}/\delta\sigma=0$, yielding the gap equation. To leading order,
$${\Sigma_0\over g^2}=-{Z_\psi\over N_f}<\bar\psi_i\psi_i>
={\rm tr}\int_p{1\over{ip{\!\!\! /}\,+\Sigma_0}},\eqno(2.4)$$
which immediately leads to an expression for $g$ in terms of $\Sigma_0$ and the
cutoff $\Lambda$:
$${1\over g^2}={8\over{(4\pi)^{d\over2}(d-2)}}\left[{{\Lambda^{d-2}}\over
{\Gamma({d\over2})}}-\Sigma_0^{d-2}
\Gamma(2-{\textstyle {d\over2}})\right].\eqno(2.5)$$
Details of how $\int_p$ is defined for $d\in(2,4)$ are given in an appendix.
The limit $\Sigma_0\to0$ of expression (2.5) defines the critical
coupling $g_c$: for $d=3$ we obtain $g_c=2\Lambda/\pi^2$. To
the same leading order the scalar propagator
$D_\sigma$ is now given by the sum over all numbers of
chained fermion - antifermion bubbles $\Pi(k^2)$:
$$\eqalign{D_\sigma^{-1}(k^2)&=Z_\sigma-\Pi(k^2)\cr
&=Z_\sigma g^2{\rm tr}\int_q{1\over{iq{\!\!\! /}\,+\Sigma_0}}
\left({1\over \Sigma_0}+{1\over{i(k{\!\!\! /}\,+q{\!\!\!
/}\,)+\Sigma_0}}\right)
\cr&=Z_\sigma g^2 {{2^{5-d}\Gamma(2-{d\over2})}
\over{(4\pi)^{d\over2}}}(k^2+4\Sigma_0^2)^{{d\over2}-1}
\int_0^{(1+4\Sigma_0^2/k^2)^{-1}}dx x^{-{1\over2}}(1-x)^{{d\over2}-2},\cr}
\eqno(2.6)$$
where use has been made of relation (2.4). The integral over $x$ in (2.6)
defines an incomplete Beta function, which in turn may be expressed in terms of
a hypergeometric function. The result is
$$D_\sigma^{-1}(k^2)=
Z_\sigma g^2{{2\Gamma(2-{d\over2})}
\over{(4\pi)^{d\over2}}}{(k^2+4\Sigma_0^2)\over\Sigma_0^{4-d}}
F(1,2-{\textstyle{d\over2}};{\textstyle{3\over2}};
-{k^2\over{4\Sigma_0^2}}).\eqno(2.7)$$
For $d=3$ we recover the usual form [9]
$$D_\sigma^{-1}(k^2)=Z_\sigma g^2{{(k^2+4\Sigma_0^2)
\tan^{-1}(\sqrt{k^2}/2\Sigma_0)}\over{2\pi\sqrt{k^2}}}.\eqno(2.8)$$
Notice that for $k^2\ll\Sigma_0^2$ $D_\sigma$ resembles the canonical form
for a boson field: $(k^2+4\Sigma_0^2)^{-1}$; indeed, in the chiral limit
it has poles at $k^2=-4\Sigma_0^2$. However, for $k^2\gg\Sigma_0^2$,
the relevant limit for the calculation of $O(1/N_f)$ loop corrections, the
asymptotic form of (2.7) is quite different:
$$\displaystyle\lim_{k^2\to\infty}{D_\sigma(k^2)}
={1\over{Z_\sigma g^2}}\left({A_d\over{\mathstrut(k^2)^{{d\over2}-1}}}
\right),\eqno(2.9a)$$
where
$$A_d={{(4\pi)^{d\over2}}\over{4\Gamma(2-{d\over2})
B({d\over2},{d\over2}-1)}}.\eqno(2.9b)$$

To renormalize the model at leading order, we may take $Z_\psi=1$, and the
physical mass $M$ to be equal to $\Sigma_0$. At higher orders
this will not be the case. Finally, to render $D_\sigma^{-1}$ finite we define
$$Z_\sigma g^2 = {1\over M^{d-2}}.\eqno(2.10)$$
The exact value of the right hand side is unimportant provided it is finite,
since it cancels from all calculations of physical quantities.
In effect we have specified $\sigma$ to have dimension
$[{\textstyle{d\over2}}]$, but since it is an auxiliary field this is purely
a matter of convention.

The simplest corrections at $O(1/N_f)$ are one loop corrections to $S_F$ and
$\Gamma_{\sigma\bar\psi\psi}$, as shown in Fig. 1 (in fact, there is a two
loop contribution to $\Gamma_{\sigma\bar\psi\psi}$ at this order [9]: since,
however, it involves a closed fermion loop with an odd number of legs, it is
finite and can be ignored). The fermion self-energy $\Sigma(p)$ is given by
$$\Sigma(p)=\Sigma_0-{g^2\over N_f}Z_\psi^2Z_\sigma\times{1\over Z_\psi}\times
\int_q{1\over{i(p{\!\!\! /}\,+q{\!\!\! /}\,)+\Sigma_0}}
D_\sigma(q^2).\eqno(2.11)$$
The integral over $q$ is logarithmically divergent but straightforward,
and leads to the result for the full inverse propagator
$$S_F^{-1}=Z_\psi\left[ip{\!\!\! /}\,\left(1+{(d-2)\over2d}{C_d\over N_f}
\ln{\Lambda\over M}\right)+\Sigma_0\left(1-{C_d\over2N_f}
\ln{\Lambda\over M}\right)\right],\eqno(2.12)$$
where we have used the fact that $M=\Sigma_0$ to leading order, and
 the numerical constant $C_d$, which we will find to be ubiquitous, is
given by
$$C_d={1\over{\Gamma(2-{d\over2})\Gamma({d\over2})B({d\over2},{d\over2}-1)}}
.\eqno(2.13)$$
In $d=3$, $C_d=4/\pi^2$. For finiteness we thus require
$$Z_\psi=1-{(d-2)\over2d}{C_d\over N_f}\ln{\Lambda\over M}\eqno(2.14)$$
and
$$\Sigma_0\equiv Z_M M=M\left(1+{(d-1)\over d}{C_d\over N_f}\ln{\Lambda\over M}
\right).\eqno(2.15)$$
Similarly, at vanishing external momentum we find the vertex to $O(1/N_f)$:
$$\Gamma_{\sigma\bar\psi\psi}=-{g\over\sqrt{N_f}}Z_\psi Z_\sigma^{1\over2}
\left(1-{C_d\over2N_f}\ln{\Lambda\over M}\right).\eqno(2.16)$$
The requirement that this expression be finite gives a condition on the
combination $Z_\sigma g^2$:
$$ gZ_\sigma^{1\over2}\propto
Z_\psi^{-1}\left(1+{C_d\over2N_f}\ln{\Lambda\over M}\right);$$
ie.
$$Z_\sigma g^2={1\over M^{d-2}}
\left(1+{{2(d-1)}\over d}{C_d\over N_f}\ln{\Lambda\over
M}\right).\eqno(2.17)$$
As we saw in Eq.(2.7), there is a second condition on $Z_\sigma g^2$ at
leading order from the requirement that the scalar propagator be finite. This
continues to be the case at higher orders, as we shall see:
consistency of the two conditions at each order
is vital for the renormalizability of the model.

To fully specify the renormalization constants to $O(1/N_f)$ we must calculate
$g$ as a function of physical mass $M$ using the two-loop gap equation (Fig.
2). The $O(1/N_f)$ contribution is:
$$\eqalign{{1\over g^2}-{1\over g^{(0)2}}&=
            -{4\over N_f}\int_p\int_q{{(3p^2-2p.q-\Sigma_0^2)}\over
{(p^2+\Sigma_0^2)^2((p-q)^2+\Sigma_0^2)}}\times D_\sigma(q^2)\cr
&=-{{        4(d-1)\Lambda^{d-2}}\over{N_f(4\pi)^{d\over2}\Gamma({d\over2})
(d-2)}}-{{2^{d-1}\Sigma_0^{d-2}\Gamma(1-{d\over2})(d-1)}\over(4\pi)^{d\over2}
},\cr}\eqno(2.18)$$
using expression (2.9) for $D_\sigma$.
The full expression for $g$ is thus:
$${1\over g^2}=
{{8\Lambda^{d-2}}\over{(4\pi)^{d\over2}\Gamma({d\over2})(d-2)}}
\left(1-{(d-1)\over2N_f}\right)
-{{8\Sigma_0^{d-2}}\over{(4\pi)^{d\over2}(d-2)}}\Gamma(2-{\textstyle{d\over2}})
\left(1-{{2^{d-3}(d-1)}\over N_f}             \right).\eqno(2.19)$$
Once again, we extract the critical coupling in the limit $\Sigma_0\to0$:
$${1\over
g_c^2}={{8\Lambda^{d-2}}\over{(4\pi)^{d\over2}\Gamma({d\over2})(d-2)}}
\left(1-{(d-1)\over2N_f}\right).\eqno(2.20)$$

We have now used up all our freedom in adjusting the constants $g$, $Z_\psi$
and $Z_\sigma$ by absorbing divergences in the fermion self-energy and vertex
corrections, and by relating quantities to a physical scale via the gap
equation. Power counting shows, however, that we have not yet exhausted the
list of divergent graphs at $O(1/N_f)$: there remain the two-loop contributions
to the scalar propagator shown in Fig. 3. If the model is renormalizable,
then the $O(1/N_f)$ expresssion for $Z_\sigma g^2$ already derived in (2.17)
together with the definition of $g$ via (2.19) must suffice to cancel the UV
divergences in $D_\sigma$. We will show this by explicit calculation in the
symmetric phase, where the massless nature of the fermion makes the loop
integrals much simpler. Of course, the UV divergence structure of a field
theory should be insensitive to the particular phase in which it resides. We
begin by following the renormalization convention of [10]: the leading order
expression for the inverse scalar propagator in the symmetric phase is now
(Cf. (2.6))
$$D_\sigma^{-1}(k^2)=Z_\sigma g^2\left[{1\over g^2}+{\rm tr}\int_q
{1\over{iq{\!\!\! /}\,}}{1\over{i(q{\!\!\! /}\,+k{\!\!\! /}\,)}}\right].
\eqno(2.21)$$
The integral over $q$ is easily done, to yield
$$D_\sigma^{-1}(k^2)=Z_\sigma g^2\left[{1\over g^2}-{{8\Lambda^{d-2}}\over
{(4\pi)^{d\over2}(d-2)\Gamma({d\over2})}}+{{(k^2)^{{d\over2}-1}}\over A_d}
\right],\eqno(2.22)$$
where $A_d$ is given in (2.9). We now impose the off-shell renormalization
condition
$$D_\sigma^{-1}(\mu^2)\equiv{2\over A_d},\eqno(2.23)$$
where $\mu$ is some arbitrary scale parameter. The power-law divergence in
(2.22)
can now be removed, first by specifying
$$Z_\sigma g^2=\mu^{2-d};\eqno(2.24)$$
then, substituting (2.22) into (2.23) we find
$${1\over g^2}={{8\Lambda^{d-2}}\over{(4\pi)^{d\over2}(d-2)\Gamma({d\over2})}}
+{{\mu^{d-2}}\over A_d}.\eqno(2.25)$$
With $Z_\sigma$ and $g^2$ both fixed, $D_\sigma^{-1}$ is completely finite:
$$D_\sigma^{-1}(k^2)={1\over{A_d\mu^{d-2}}}\left((k^2)^{{d\over2}-1}+\mu^{d-2}
\right).\eqno(2.26)$$
We see that in the symmetric phase $D_\sigma(k^2)$ has no poles on the physical
sheet, and so $\mu$ cannot be interpreted as a scalar mass: indeed the $\sigma$
field in this case does not interpolate a stable particle. Instead, $\mu$ can
be regarded as the width of an unstable resonance, but as shown later, it can
still serve as an inverse correlation length in the statistical mechanics
sense.
The critical coupling $g_c$ is obtained in the limit $\mu\to0$, whereupon we
find the same result as for the $\Sigma_0\to0$ limit of (2.5).
Note that in (2.25)
$g<g_c$, which is consistent with being in the symmetric phase. To leading
order
no further divergences are present, so as before $Z_\psi=1$.

It is important to note that the asymptotic form of (2.26) matches that of the
propagator in the broken phase (2.9), as it must do. It is therefore no
surprise that we obtain almost identical results for the fermion wavefunction
and vertex corrections at $O(1/N_f)$ as for the broken phase (Fig. 1), the
only difference being that now no mass renormalization is required, due to the
chiral symmetry of the model. We obtain two conditions for the renormalization
constants essentially identical to Eqs.(2.14,17):
$$\eqalignno{Z_\psi&=1-{(d-2)\over 2d}{C_d\over N_f}\ln{\Lambda\over\mu};
&(2.27)\cr
Z_\sigma g^2&={1\over\mu^{d-2}}\left(1+{{2(d-1)}\over d}{C_d\over N_f}
\ln{\Lambda\over\mu}\right),&(2.28)\cr}$$
with $C_d$ given by (2.13).

Since there are no non-trivial solutions to the
gap equation for $g<g_c$, the renormalization of the model can only be
completed
by extracting $g(\Lambda,\mu)$ from the full inverse scalar propagator to
$O(1/N_f)$, which necessitates a two-loop calculation. The relevant
contributions are shown diagramatically in Fig. 3  (once again, there is
a finite three-loop contribution which we ignore [9]). For external momentum
$k$, the two-loop contribution to $D_\sigma^{-1}(k^2)$ is
$$\eqalign{\Pi^{(1)}(k^2)=
           -{{Z_\sigma g^2}\over N_f}&\int_q
{A_d\over{(q^2)^{{d\over2}-1}+\mu^{d-2}}}\cr
&\times\int_p\left\{{\rm tr}\left({1\over{ip{\!\!\! /}\,}}{1\over{i(p{\!\!\! /}
\,+k{\!\!\! /}\,)}}{1\over{i(p{\!\!\! /}\,+k{\!\!\! /}\,+q{\!\!\! /}\,)}}
{1\over{i(p{\!\!\! /}\,+q{\!\!\! /}\,)}}\right)+2{\rm tr}\left(
{1\over{ip{\!\!\! /}\,}}{1\over{i(p{\!\!\! /}\,+q{\!\!\! /}\,)}}
{1\over{ip{\!\!\! /}\,}}{1\over{i(p{\!\!\! /}\,+k{\!\!\!
/}\,)}}\right)\right\}.
\cr}\eqno(2.29)$$
We will spare the reader the full details of the calculation, and merely
remark that the integral over $q$ is considerably simplified by our choice of
momentum routing through the internal scalar line. The divergences are
generically of two kinds: a power-law form
$$(i)\;\;\;\sim\;{{\Lambda^{d-2}}\over(d-2)}-\mu^{d-2}\ln{\Lambda\over\mu};
\eqno(2.30) $$
and a momentum-dependent logarithm
$$(ii)\;\;\;\sim\;\;(k^2)^{{d\over2}-1}\ln{\Lambda\over{\{\mu\vert k\}}},
\eqno(2.31) $$
where the symbol $\{\mu\vert k\}$ denotes a function of both $\mu$ and $k$
which simplifies in three limits of interest:
$$\eqalign{&\displaystyle\lim_{\mu\to0}\{\mu\mid k\}=\sqrt{k^2};\cr
           &\displaystyle\lim_{k  \to0}\{\mu\mid k\}=\mu;\cr
           &\displaystyle\lim_{\mu^2=k^2}\{\mu\mid k\}=\mu.\cr}\eqno(2.32) $$
Further details on how forms $(i)$ and $(ii)$ are extracted from (2.29) are
given in the appendix.

After some work we arrive at the full expression for $D_\sigma^{-1}(k^2)$ to
$O(1/N_f)$:
$$\eqalign{D_\sigma^{-1}(k^2)&=Z_\sigma g^2\Biggl[{1\over g^2}-
{{8\Lambda^{d-2}}\over{(4\pi)^{d\over2}(d-2)\Gamma({d\over2})}}+
{{(k^2)^{{d\over2}-1}}\over A_d}\cr
&\phantom{=Z_\sigma}+{{8B({d\over2}-1,{d\over2}-1)
\Gamma(2-{d\over2})}\over{(4\pi)^d\Gamma({d\over2})}}{A_d\over N_f}(d-1)\left\{
\left({\Lambda^{d-2}\over(d-2)}-\mu^{d-2}\ln{\Lambda\over\mu}\right)
-{2\over d}(k^2)^{{d\over2}-1}\ln{\Lambda\over{\{\mu\vert k\}}}\right\}
\Biggr]\cr
&\phantom{=1}\cr
&=Z_\sigma g^2\left[{1\over g^2}-{1\over g_c^2}-(d-1){{\mu^{d-2}}\over A_d}
{C_d\over N_f}\ln{\Lambda\over\mu}+{{(k^2)^{{d\over2}-1}}\over A_d}\left(
1-{{2(d-1)}\over d}{C_d\over N_f}\ln{\Lambda\over{\{\mu\vert k\}}}\right)
\right],\cr}\eqno(2.33) $$
where $g_c$ is identical to the result for the broken phase, Eq.(2.20).
Once again, we renormalize using condition (2.23), plus the expression (2.28)
for $Z_\sigma g^2$ coming from vertex renormalization. We find an expression
for $g$:
$${1\over g^2}={1\over g_c^2}+{{\mu^{d-2}}\over A_d}\left(1+
{{(d-1)(d-2)}\over d}{C_d\over N_f}\ln{\Lambda\over\mu}\right).\eqno(2.34) $$
Substituting back into (2.33),  we find the renormalized inverse scalar
propagator:
$$\eqalign{D_\sigma^{-1}(k^2)&=Z_\sigma g^2\left[{{\mu^{d-2}}\over A_d}\left(
1-{{2(d-1)}\over d}{C_d\over N_f}\ln{\Lambda\over\mu}\right)+
{{(k^2)^{{d\over2}-1}}\over A_d}\left(
1-{{2(d-1)}\over d}{C_d\over N_f}\ln{\Lambda\over{\{\mu\vert
k\}}}\right)\right]
\cr
&\phantom{=1}\cr
&={1\over{A_d\mu^{d-2}}}\left[\mu^{d-2}+(k^2)^{{d\over2}-1}\left(
1+{{2(d-1)}\over d}{C_d\over N_f}\ln{{\{\mu\vert k\}}\over\mu}\right)\right].
\cr}\eqno(2.35) $$
This expression is manifestly UV finite both for $k^2=0$, and in the critical
limit $\mu\to0$, although in the latter case there is the usual IR divergence
associated with massless particles.

Since we have been able to adjust $Z_\psi$, $Z_\sigma$ and $g$ to eliminate
all divergences in Green functions to $O(1/N_f)$, at least in the symmetric
phase, we have succeeded in our aim of showing the model is renormalizable to
this order. In effect the fact that the combination $Z_\sigma g^2$ serves both
to correct the fermion-scalar vertex and the inverse scalar propagator is
really
a statement that the renormalized four-point fermion Green function is finite
to this order (Fig. 4).  If we recall that the underlying model we started
from was one with an explicit four-fermi interaction, we can now understand why
$Z_\sigma g^2$ is apparently overdetermined: in the four-fermi model there can
only be {\it two\/} adjustable parameters: the unphysical $Z_\psi$; and the
strength of the four-point coupling. Another consequence is that for
$k^2\gg M^2,\mu^2$, logarithmic contributions to the four-point scattering
of the form $\ln(k/M)$ must also cancel, which means that in this asymptotic
limit the renormalized four-fermion scattering amplitude assumes a universal
form $A_d/N_f k^{d-2}$. This receives no $1/N_f$ corrections if the model is
renormalizable.

We now consider the model from the viewpoint of statistical mechanics,
by calculating the critical exponents associated with the chiral symmetry
breaking transition, and demonstrating a relation between renormalizability
and hyperscaling.
All essential information about a continuous phase transition is encoded in its
critical exponents, and it is straightforward to calculate these to
$O(1/N_f)$ from the gap equation. The order parameter is of course
the condensate
$<\bar\psi\psi>$, but from (2.4) we can equally well consider
$\Sigma_0$, since it shares the same non-analytic behavior as $g\to g_{c+}$,
$\Lambda/M\to\infty$. Parameterizing the distance from criticality by
$t\equiv g^2( g_c^{-2}-g^{-2})$, we find from (2.19) the critical exponent
$\beta$:
$$t\propto \Sigma_0^{\textstyle  {1\over\beta}}\propto \Sigma_0^{d-2},
\eqno(2.36)$$
whence
$$\beta={1\over(d-2)}+O\left({1\over N_f^2}\right).\eqno(2.37)$$

To determine the other critical exponents describing the critical behavior of
the order parameter, we need to consider the effects of introducing an explicit
bare mass term $Z_\psi m\bar\psi_i\psi_i$ into the Lagrangian (2.1).
To leading order $M=\Sigma_0+m$, and
we obtain an inverse scalar propagator which now has a
a contribution diverging as a power of $\Lambda$ at leading order:
$$\eqalign{D_\sigma^{-1}(k^2;m  )&=Z_\sigma g^2{\rm tr}\int_q
{1\over{iq{\!\!\! /}\,+M}}\left({1\over \Sigma_0}+
{1\over{i(k{\!\!\! /}\,+q{\!\!\! /}\,)+M}}\right)\cr
&=Z_\sigma g^2\Biggl\{{{2\Gamma(2-{d\over2})}
\over{(4\pi)^{d\over2}}}{(k^2+4M^2)       \over M^{4-d}}
F(1,2-{\textstyle{d\over2}};{\textstyle{3\over2}};-{k^2\over{4M^2}})\cr
&\phantom{ZZZZZZZ}
+{8\over{(4\pi)^{d\over2}(d-2)}}{m  \over \Sigma_0}\left[{\Lambda^{d-2}\over
\Gamma({d\over2})}-M^{d-2}\Gamma(2-{\textstyle{d\over2}})\right]\Biggr\};\cr}
\eqno(2.38a)$$
$$\displaystyle\lim_{k^2\to\infty}D_\sigma^{-1}(k^2;m)=
Z_\sigma g^2\left[{{(k^2)^{{d\over2}-1}}\over A_d}+
{8\over{(4\pi)^{d\over2}(d-2)}}{m  \over \Sigma_0}\left({{\Lambda^{d-2}}\over
\Gamma({d\over2})}-M^{d-2}\Gamma(2-{\textstyle{d\over2}})\right)\right].
\eqno(2.38b)
$$
It is interesting to note that away from the chiral limit $D_\sigma$ no longer
has poles in the complex $k$-plane -- therefore to define a ``$\sigma$ mass''
we are forced to use the definition (1.10,22).
If we substitute the asymptotic form (2.38b) into the two-loop gap equation
(2.18), with $M$ replacing $\Sigma_0$ in the fermi propagators,
we obtain the full gap equation in the presence of external masss:
$${\Sigma_0\over g^2}={M\over g_c^2}-M\left[{{8M^{d-2}\Gamma(2-{d\over2})}\over
{(4\pi)^{d\over2}(d-2)}}-{{8(d-1)}\over{(4\pi)^{d\over2}(d-2)^2
\Gamma({d\over2})}}{C_d\over N_f}{{\Lambda^{d-2}m  }\over \Sigma_0}
\ln\left(1+{{(d-2)C_d}\over2}{\Sigma_0\over m  }\right)\right].\eqno(2.39)$$
Setting $g=g_c$ and taking the limit $m\ll\Sigma_0\sim M$ we find
$${m  \over g_c^2}\left[1+{(d-1)\over(d-2)}
{C_d\over N_f}\ln{\Sigma_0\over m  }\right]
={{8\Sigma_0^{d-1}\Gamma(2-{d\over2})}\over{(4\pi)^{d\over2}(d-2)}};\eqno(2.40)
$$
ie.
$$\delta=(d-1)\left[1+{C_d\over N_f}\right],\eqno(2.41)$$
where we have assumed that the logarithm exponentiates
beyond $O(1/N_f)$.
To extract $\gamma$, we must return to the full expression (2.38a) for
$D_\sigma(k^2;m)$, substitute it into (2.18) and evaluate
$\partial/\partial m\vert_{m=0}$ on the resulting gap equation. After some
work we find
$${1\over g^2}\left(1+(d-1){C_d\over N_f}\ln{\Lambda\over\Sigma_0}
\right)=(1+\chi)\Sigma_0^{d-2}{{8\Gamma(2-{d\over2})}\over
(4\pi)^{d\over2}}.\eqno(2.42)$$
Near criticality $\chi\gg1$, from which we can infer
$$\chi\propto \Sigma_0^{-[d-2+(d-1)C_d/N_f]}
\propto t^{-[1+ (d-1)C_d/(d-2)N_f]}\eqno(2.43)$$
using (2.36), ie:
$$\gamma=1+{{ (d-1)}\over{(d-2)}}{C_d\over N_f}.\eqno(2.44)$$

We now see that our derived values for $\beta$, $\delta$ and $\gamma$ are
consistent to $O(1/N_f)$ with the scaling relation
$$\gamma=\beta(\delta-1),\eqno(2.45)$$
which raises the issue of whether the exponents calculated in this
way obey scaling or hyperscaling relations. To develop this
picture we need to calculate further critical exponents not directly
determined by the behavior of the order parameter, and hence need to
supply some physical insight. The
obvious choice for correlation length is the inverse physical fermion
mass $M^{-1}$. Using Eqs. (2.15) and (2.36) we find
$$t\propto\xi^{\textstyle  {-{1\over\nu}}}\propto M^{(d-2)
[1-(d-1)C_d/dN_f]},\eqno(2.46)$$
leading to a prediction for the exponent $\nu$:
$$\nu={1\over(d-2)}\left[1+{{(d-1) }  \over d     } {C_d\over N_f}\right].
\eqno(2.47)$$

We can calculate the final exponent $\eta$ in two different ways. Firstly
and more directly, consider the renormalized scalar propagator (ie. two-point
correlator) (2.35): at criticality we expect the scaling form
$$D_\sigma^{-1}\propto k^{2-\eta}.\eqno(2.48)$$
{}From the $\mu\to0$ limit of (2.35) we deduce
$$\eta=4-d-{{2(d-1)}\over d}{C_d\over N_f}.\eqno(2.49)$$
Of course, strictly we should consider correlations of the scalar composite
$\bar\psi\psi$, which defines the true order parameter. Following the
treatment of [10], we can renormalize this local operator, and use (2.4)
and (2.15), to write
$$M=-{{Z_\psi Z_M^{-1}g^2}\over N_f}<\bar\psi_i\psi_i>=
-{{Z_\psi Z_M^{-1}g^2}\over N_f}Z_{\bar\psi\psi}<(\bar\psi_i\psi_i)_R>,
\eqno(2.50)$$
where the constant $Z_{\bar\psi\psi}$ has been introduced to
renormalize $\bar\psi_i\psi_i$. Now, both sides of (2.50)
are cutoff-independent (since  $M$ is a physical mass), as is
$(\bar\psi_i\psi_i)_R$ itself; note also that a factor of $Z_\psi$ is absorbed
in taking the trace to calculate $<\bar\psi\psi>$. We
conclude that
$$\Lambda{{\partial\>}\over{\partial\Lambda}}
\left(Z_M^{-1}g^2Z_{\bar\psi\psi}\right)=0,
\eqno(2.51)$$
ie.
$$\tilde\gamma_{\bar\psi\psi}\equiv{\Lambda\over{Z_{\bar\psi\psi}}}
{{\partial\>}\over{\partial\Lambda}}Z_{\bar\psi\psi}
 ={{g^2\Lambda}\over Z_M}
{{\partial\>}\over{\partial\Lambda}}\left({Z_M\over g^2}\right)
 =(d-2){g^2\over g_c^2}+{(d-1)\over d}{C_d\over N_f},    \eqno(2.52)$$
where we have used (2.5,15), and the quantity denoted by
$\tilde\gamma_{\bar\psi\psi}$ is the {\it anomalous dimension\/} of the
operator $\bar\psi\psi$ defined in the conventional field-theoretic sense [18].
At the fixed point $g=g_c$, we can use the resulting
\hbox{$\tilde\gamma_{\bar\psi\psi}=(d-2)+(d-1)C_d/dN_f$} to derive $\eta$.
We expect the
renormalized two-point correlation function
$<(\bar\psi\psi(0))_R(\bar\psi\psi(x))_R>$ to
scale asymptotically in momentum space as
$$\Gamma_R^{(0;2)}(k^2)\propto k^{-[d-2d_{\bar\psi\psi}^{(0)}+
2\tilde\gamma_{\bar\psi\psi}]}\propto k^{-[2-d+2\tilde\gamma_{\bar\psi\psi}]},
\eqno(2.53)$$
where we have introduced the canonical (ie. free-field) scaling dimension for
$\bar\psi\psi$:
$$d_{\bar\psi\psi}^{(0)}=d-1.\eqno(2.54)$$
or in other words $\eta=d-2\tilde\gamma_{\bar\psi\psi}$ using (2.48).
The value for $\eta$ thus derived is identical to (2.49), demonstrating that
these rather formal arguments about operator renormalization are valid, and
that for all practical purposes $\sigma$ and $\bar\psi\psi$ can be considered
identical.

The calculation of the critical exponents is now complete:
to $O(1/N_f)$ the values we have computed for $\beta$,
$\delta$, $\gamma$, $\nu$ and $\eta$ satisfy the hyperscaling relations (1.17):
$$2\beta+\gamma=d\nu;\;\;2\beta\delta-\gamma=d\nu;\;\;\beta={1\over2}\nu
(d-2+\eta);\;\;\gamma=\nu(2-\eta).\eqno(2.55)$$
In the realistic case $d=3$ discussed in the rest of this paper, the values
coincide with those given in Eq.(1.3).
It is also interesting to examine the limits $d\to4_-$ and $d\to2_+$. Noting
that $C_d$ vanishes in each case, as $(4-d)$ and as $(d-2)$ respectively,
we find for $d\to4$ we recover mean-field exponents;
$$\beta={\textstyle{1\over2}};\;\;\delta=3;\;\;\gamma=1;\;\;
\nu={\textstyle{1\over2}};\;\;\eta=0,\eqno(2.56)$$
while for $d\to2$ we find
$$\beta={1\over{(d-2)}};\;\;\delta=1;\;\;
\gamma=1+{1\over2N_f};\;\;
\nu={1\over{(d-2)}};\;\;\eta=2.\eqno(2.57)$$

It is interesting to note that we could have inferred the renormalizability of
the model merely by considering the exponents, without performing the integral
(2.29). Suppose
we had {\it assumed\/} hyperscaling to derive $\eta$ from the gap equation
exponents $\beta$, $\delta$, and $\gamma$, and made no further assumptions
about
operator renormalization, then we could have
used relations (2.48) to {\it predict\/} the divergence structure of the
scalar propagator at criticality:
$$D_\sigma^{-1}(k^2)\vert_{g=g_c}\propto k^{2-\eta}\sim k^{d-2}\left[
1-{{2(d-1)}\over d}{C_d\over N_f}\ln{\Lambda\over k}\right].\eqno(2.58)$$
The argument of the logarithm is determined on dimensional grounds, since the
cutoff is the only other scale left in the problem. Similarly we could use an
alternative definition of susceptibility, together with the derived values of
$\gamma$ and $\nu$, to deduce the momentum-independent divergence:
$$\chi^{-1}\equiv D_\sigma^{-1}(k^2=0)\propto t^\gamma\propto
\xi^{-{\textstyle{\gamma\over\nu}}}\sim M^{d-2}\left[1-{{2(d-1)}\over d}
{C_d\over N_f}\ln{\Lambda\over M}\right].\eqno(2.59)$$
In both cases we can check that $Z_\sigma g^2$ given in (2.17)  eliminates all
dependence on the cutoff $\Lambda$ and renders $D_\sigma^{-1}$ finite. Hence
we can infer renormalizability at this order
without ever doing a two-loop calculation
(although the gap equation has two loops, it has one fewer fermi propagator,
and
no external momentum dependence, and so is much simpler!).

There would appear to be an intimate connection between hyperscaling and
renormalizability which we can demonstrate to all orders in $1/N_f$
without doing any detailed calculations. Consider the effect of shifting
the $\sigma$ field in the Lagrangian (2.1) so that the fermions pick up
a mass term $Z_\psi\Sigma_0\bar\psi_i\psi_i$. It is then
straightforward to derive the following functional identity [19]:
$${\partial\over{\partial\Sigma_0}}{{\delta^2\Gamma}\over{\delta\bar\psi\delta
\psi}}={{\sqrt{N_f}}\over{gZ_\sigma^{1\over2}}}{{\delta^3\Gamma}\over
{\delta\bar\psi\delta\psi\delta\sigma^\prime}},\eqno(2.60)$$
where $\sigma^\prime$ is the shifted scalar field and
$\Gamma=\Gamma[\psi,\bar\psi,\sigma]$ is the  effective action.
We can use this relation in the
broken phase    to relate the fermion
self-energy $\Sigma(k^2)$ to the full fermion-scalar vertex [9]:
$$\Sigma(0)=-\Sigma_0{\sqrt{N_f}\over{gZ_\sigma^{1\over2}Z_\psi}}
\Gamma_{\sigma\bar\psi\psi}(0),\eqno(2.61)$$
The chiral symmetry of the original Lagrangian implies
$\Sigma(0)$ is proportional to $\Sigma_0$, and power counting shows that
$\Gamma_{\sigma\bar\psi\psi}$ and hence $\Sigma$ are at most
logarithmically divergent functions of $\Lambda$.
Using the definitions implicit in (2.11-16),    we deduce
$$Z_\psi^{-1}Z_M^{-1}\Sigma_0\propto
           -{\Sigma_0\over g}Z_\sigma^{-{1\over2}}Z_\psi^{-1},$$
ie.
$$Z_M^2\propto Z_\sigma g^2.\eqno(2.62)$$
The constant of proportionality is $O(M^{2-d})$ from (2.10).
Now, we know that $Z_M$ has the form
$$Z_M=    1+a\ln{\Lambda\over M},\eqno(2.63)$$
where we assume $a     \sim O(1/N_f)$ is small, so that
$\Sigma_0\propto M^{1-a}$.      By definition, we also have
$t\propto \Sigma_0^{1\over\beta}\propto M_{\phantom{0}}^{1\over\nu}$,
which enables us to relate $a$ to $\beta$ and $\nu$ assuming that higher powers
of the logarithm  exponentiate      as required by the hypothesis of
power-law scaling near a critical point:
$$1-a     ={\beta\over\nu},\eqno(2.64)$$
so from (2.63,64)  we deduce
$$Z_\sigma g^2\propto1+2\left(1-{\beta\over\nu}\right)\ln{\Lambda\over M}.
\eqno(2.65)$$
{}From previous we know that if the model is renormalizable $Z_\sigma g^2$ must
suffice to render the inverse scalar propagator $D_\sigma^{-1}$ finite, by
cancelling both $k$-dependent and $k$-independent logarithmic divergences,
once power-law divergences have been removed by tuning $g^2$ to its
critical value (Cf. Eq.(2.33)).
Consider each case separately. At criticality, we must have on dimensional
grounds
$$D_\sigma^{-1}(k^2)\propto k^{2-\eta}\sim k^{d-2}\left[1-(4-d-\eta)
\ln{\Lambda\over k}\right].\eqno(2.66)$$
If the logarithmic divergence in (2.66) is to be cancelled by the $Z_\sigma
g^2$
of (2.65),  then
$$2\left(1-{\beta\over\nu}\right)=4-d-\eta,$$
ie.
$$\beta={1\over2}\nu(d-2+\eta),\eqno(2.67)$$
which is one of the hyperscaling relations (2.55). Similarly in the limit
$k^2\to0$ we have
$$D_\sigma^{-1}(0)\propto M^{\textstyle{\gamma\over\nu}}\sim
M^{d-2}\left[1-\left(2-d+{\gamma\over\nu}\right)\ln{\Lambda\over M}\right],
\eqno(2.68)$$
so that renormalizability implies
$$2\beta+\gamma=d\nu,\eqno(2.69)$$
which is another hyperscaling relation. Hence with a few simple assumptions
about the form of divergences at $O(1/N_f)$ and beyond,
we can see the equivalence of
hyperscaling and renormalizability. This puts the renormalizability of the
model on a very physical footing; indeed, the main hypothesis behind
both properties is the assumption that there is only one important
physical length scale (correlation length) whose divergence in cutoff units
at the critical point controls the scaling of all other physical quantities.
In this model the physical length scale is the renormalized fermion mass $M$
in the broken phase and the scalar width $\mu$ in the symmetric phase: both
ratios $\Lambda/M$ and $\Lambda/\mu$ diverge with the same critical exponent
$\nu$ in the chiral limit at the phase transition, which defines the
continuum limit. The
continuity of $\nu$ is demanded on physical grounds [20], since otherwise
the free energy of the system would be discontinuous across the phase
transition.

Suppose we had found that in
the broken phase the value of $\nu$ demanded by hyperscaling differed from the
value $\nu_M$ derived by assuming that the renormalized fermion mass $M$
is indeed an inverse correlation length. We would then have been
forced to conclude that either the interaction strength tended to zero as
$g\to g_c$ ($\nu_M>\nu$), so that the model described free massive fermions,
or that the fermions became very massive and decoupled ($\nu_M<\nu$), leaving
a theory of non-interacting scalar bound states. In either case the model would
have had a trivial continuum limit. However, this scenario is impossible in the
present model, since we know from the identity (2.60,61)  that the
definition of renormalized fermion mass is tied to the renormalization constant
$Z_\sigma g^2$, which in turn cancels out divergences in the inverse scalar
propagator if and only if hyperscaling is obeyed. Therefore a model in which
$\nu_M\not=\nu$ would be non-renormalizable. Thus in this model, and as
we shall see, in similar four-fermi models, the issues of hyperscaling,
renormalizability, and non-triviality are inseparable.

Finally, we note that in addition to references [9,11], there have been
other calculations beyond leading order in the Gross-Neveu model for
$2<d<4$. Our result for $Z_\psi$ is consistent with the calculation of Hikami
and Muta [21], modulo different definitions of the trace over the Dirac
algebra.
We also agree with the results of [9,11,22] for $d=3$. Gracey [23] has
also computed the exponent $\eta$ (in our notation) and has actually
obtained   $Z_\psi$ to $O(1/N_f^2)$: his sophisticated methods rely
on being exactly at the critical point and hence cannot be used to
calculate the other exponents. Other recent calculations beyond leading order
for general $d$ have appeared in [24].

Finally in this section, for completeness, we consider extensions of the
Gross-Neveu model to cases where the spontaneously broken symmetry is a
continuous one [5]. There are two interesting cases; the symmetries in the
Lagrangian are either ${\rm U(1)}_L\otimes{\rm U(1)}_R$
($\equiv{\rm U(1)}_V\otimes{\rm U(1)}_A$):
$${\cal L}=\bar\psi_i\partial{\!\!\! /}\,\psi_i
-{g^2\over{2N_f}}\left[(\bar\psi_i\psi_i)^2-(\bar\psi_i\gamma_5\psi_i)^2
\right];\eqno(2.70a)$$
or ${\rm SU(2)}_L\otimes{\rm SU(2)}_R$:
$${\cal L}=\bar\psi_{ip}\partial{\!\!\! /}\,\psi_{ip}
-{g^2\over{2N_f}}\left[(\bar\psi_{ip}\psi_{ip})^2-(\bar\psi_{ip}\gamma_5
\vec\tau_{pq} \psi_{iq})^2\right],\eqno(2.70b)$$
where $\vec\tau$  are the Pauli matrices, normalised to
${\rm tr}(\tau^\alpha\tau^\beta)=2\delta^{\alpha\beta}$, and we show the
indices $p,q$ running from 1 to 2 explicitly. In $2<d<4$ dimensions we
define $\gamma_5$ thus: $\gamma_5^2=1$, $\{\gamma_5,\gamma_\mu\}=0$,
${\rm tr}(\gamma_5\gamma_{\mu_1}\ldots\gamma_{\mu_n})=0$.
We can immediately bosonize the
models, and introduce renormalization constants as in (2.1):
$${\cal L}=Z_\psi\bar\psi_i\partial{\!\!\! /}\psi_i+
{g\over\sqrt{N_f}}Z_\psi Z_\phi^{1\over2}\left[\sigma\bar\psi_i\psi_i
+i\pi\bar\psi_i\gamma_5\psi_i\right]+{1\over2}Z_\phi(\sigma^2+\pi^2);
\eqno(2.71a)$$
$${\cal L}=Z_\psi\bar\psi_i\partial{\!\!\! /}\psi_i+
{g\over\sqrt{N_f}}Z_\psi Z_\phi^{1\over2}\left[\sigma\bar\psi_i\psi_i
+i\vec\pi . \bar\psi_i\gamma_5\vec\tau \psi_i\right]+{1\over2}Z_\phi
(\sigma^2+\vec\pi^2 ),\eqno(2.71b)$$
where $\sigma$ is an auxiliary scalar and $\pi$ an auxiliary pseudoscalar.
Note that in each case the combination $\phi=\sigma+i\pi$  ($\sigma+i\vec\pi $)
is proportional to an element of the chiral group, so that eg. in the
SU(2)$\otimes$SU(2) case we have $\psi_L\mapsto U\psi_L$;
$\psi_R\mapsto V\psi_R$; $\phi\mapsto V\phi U^{-1}$, where
$\psi_{L,R}=(1\pm\gamma_5)\psi/2$, and $U$ and $V$ are SU(2) matrices. This
property does not in general extend to higher flavor groups such as U($n$).
Model (2.71b) closely resembles the original Gell-Mann - L\'evy $\sigma$-model
[25]. Notice also that the same renormalization constant $Z_\phi$ serves for
both $\sigma$ and $\pi$ fields: this is the result of a Ward identity which
we discuss below.

To leading order in $1/N_f$ model (2.70a) has an identical gap equation to
(2.5) and expression for $D_\sigma$ (2.7). The only new feature at this order
is the pion propagator, which in the broken phase is given by
$$D_\pi^{-1}(k^2)=
Z_\phi   g^2{{2\Gamma(2-{d\over2})}
\over{(4\pi)^{d\over2}}}{ k^2             \over\Sigma_0^{4-d}}
F(1,2-{\textstyle{d\over2}};{\textstyle{3\over2}};
-{k^2\over{4\Sigma_0^2}}).\eqno(2.72)$$
The pole at $k^2=0$ explicitly shows that the $\pi$ field has a Goldstone
mode in the broken phase. In the symmetric phase we find as before (2.24,26):
$$D_\sigma^{-1}(k^2)=D_\pi^{-1}(k^2)={{Z_\phi g^2}\over A_d}
[(k^2)^{{d\over2}-1}+\mu^{d-2}].\eqno(2.73)$$
The SU(2)$\otimes$SU(2) model goes through in very similar fashion apart from
an overall factor of 2 in the gap equation from the trace over Pauli indices,
and a corresponding factor of ${1\over2}$ in $D_\sigma$ and $D_\pi$, the
latter of course now carrying an implicit $\delta^{\alpha\beta}$.

The calculation of $O(1/N_f)$ corrections proceeds as before, the only new
feature being the appearance of a logarithmic divergence in the $O(1/N_f)$
correction to the gap equation containing an internal $\pi$ line. The
renormalization constants are found to be, for U(1)$\otimes$U(1):
$$\eqalign{
Z_\psi&=1-{{(d-2)}\over d}{C_d\over N_f}\ln{\Lambda\over M};\cr
Z_M   &=1+{{(d-2)}\over d}{C_d\over N_f}\ln{\Lambda\over M};\cr
Z_\phi g^2&\propto1+{{2(d-2)}\over d}{C_d\over N_f}\ln{\Lambda\over M},\cr}
\eqno(2.74)$$
with $g^2$ given by the gap equation:
$${1\over g^2}={{8\Lambda^{d-2}}\over{(4\pi)^{d\over2}\Gamma({d\over2})(d-2)}}
\left[1-{{(d-2)}\over N_f}\right]-
{{8\Sigma_0^{d-2}\Gamma(2-{d\over2})}\over{(4\pi)^{d\over2}(d-2)}}
\left[1-{{2^{d-2}(d-2)}\over N_f}+(d-2){C_d\over N_f}\ln{\Lambda\over\Sigma_0}
\right].\eqno(2.75)$$
Similarly for SU(2)$\otimes$SU(2) we find:
$$\eqalign{
Z_\psi&=1-{{(d-2)}\over d}{C_d\over N_f}\ln{\Lambda\over M};\cr
Z_M   &=1+{{(d-4)}\over2d}{C_d\over N_f}\ln{\Lambda\over M};\cr
Z_\phi g^2&\propto1+{{(d-4) }\over d}{C_d\over N_f}\ln{\Lambda\over M},\cr}
\eqno(2.76)$$
and the gap equation:
$${1\over g^2}={{16\Lambda^{d-2}}\over{(4\pi)^{d\over2}\Gamma({d\over2})(d-2)}}
\left[1-{{(2d-5)}\over2N_f}\right]-
{{16\Sigma_0^{d-2}\Gamma(2-{d\over2})}\over{(4\pi)^{d\over2}(d-2)}}
\left[1-{{2^{d-2}(2d-5)}\over2N_f}+{3\over2}(d-2){C_d\over N_f}
\ln{\Lambda\over\Sigma_0}\right].\eqno(2.77)$$
Once again, we find that the combination $Z_\phi g^2$ derived via vertex
renormalization is also sufficient to cancel divergences in the two-loop
scalar and pseudoscalar propagator corrections. The calculation of critical
indices is straightforward, and the results for these, plus the critical
coupling $g_c^2$, are tabulated in Table I. It is easily checked that the
exponents satisfy the hyperscaling relations (2.55) to $O(1/N_f)$.

The arguments (2.60 - 69) showing the relation between renormalizability
and hyperscaling go through exactly as before, except that now because the
broken symmetry is continuous, we can start from a fully-fledged Ward identity,
eg. for U(1)$\otimes$U(1) we have
$$-<\sigma>\Gamma_{\pi\bar\psi\psi}(p_\pi=0)={i\over2}\{S_F^{-1},\gamma_5\}.
\eqno(2.78)$$
Using relations (2.3) and (2.12) we arrive at a result analoguous to (2.62):
$$Z_M^2\propto Z_\pi g^2.\eqno(2.79)$$
To complete the argument we need a further Ward identity:
$$\Gamma_{\pi\bar\psi\psi}+<\sigma>\Gamma_{\pi\sigma\bar\psi\psi}(p_\pi=0)=
{i\over2}\{\Gamma_{\sigma\bar\psi\psi},\gamma_5\}.\eqno(2.80)$$
The four-point function $\Gamma_{\pi\sigma\bar\psi\psi}$ is constructed from
superficially convergent graphs, so needs no renormalization. Assuming
$\gamma_5$ commutes with the scalar vertex, we see that $Z_\sigma=Z_\pi=Z_\phi$
as promised. Equivalent arguments hold for the SU(2)$\otimes$SU(2) model.

We end this section with a comment about proving renormalizability. We have
stressed the physical equivalence of renormalizability and hyperscaling for
model field theories of this kind, and, we suspect, for a variety of other
models with a non-trivial fixed point in $2<d<4$ [6,26]. A
crucial point in
our argument has been the existence of a chiral symmetry leading to identities
such as (2.62,79). The model will remain renormalizable so long as this
symmetry is only broken by soft terms, ie. a bare fermion mass. (The effects
of introducing a symmetry-breaking term $(\bar\psi_i\psi_i)^3$ into ${\cal L}$
and the consequent non-renormalizability of the model have been discussed in
[22]).
It has been
stated [9,22] that the renormalizability of the model follows from
power-counting considerations alone, once the leading order form of the
scalar propagator (2.7) has been established. We feel that this ignores the
necessity for the consistency relations between the renormalization constants
we have presented here -- hence the chiral symmetry of the model must play
a fundamental role in the full proof. Of course, this is not a new situation:
power-counting alone does not suffice to prove the renormalizability of gauge
theories.
Moreover, the SU(2)$\otimes$SU(2) model provides a specific example where
next-to-leading order corrections render scalar exchange {\it harder\/}
than leading order, since $\eta>4-d$. Power-counting arguments to analyze the
degree of divergence of the graphs would fail in this case unless the
momentum dependence of the dressed vertices is taken into account. The
demonstration that $O(1/N_f)$ and higher corrections to scaling dimensions
cancel in the correct fashion, so as to leave the set of primitively
divergent graphs given by the leading order predictions unchanged, depends on
consistency relations derived from Ward identities.
\vskip 1.5 truecm
\noindent{\bf3. Lattice Formulation of the Gross-Neveu Model}

The Gross-Neveu model in its bosonized form (1.2) may be formulated in three
dimensions on a spacetime lattice using the following action:
$$S=\sum_{i=1}^{N_f/2}\left(\sum_{x,y} \bar\chi_i(x){\cal M}_{x,y}\chi_i(y)
+{1\over8}\sum_x\bar\chi_i(x)\chi_i(x)\sum_{<{\tilde x},x>}\sigma({\tilde x})
\right)+{N_f\over{4g^2}}\sum_{\tilde x}\sigma^2({\tilde x}),\eqno(3.1)$$
where $\chi_i,\bar\chi_i$ are Grassmann-valued staggered fermion fields
defined on the lattice sites, the auxiliary scalar field $\sigma$ is defined
on the dual lattice sites, and the symbol $<{\tilde x},x>$ denotes the set of 8
dual lattice sites ${\tilde x}$ surrounding the direct lattice site $x$.
The lattice spacing $a$ has been set to one for convenience. The
fermion kinetic operator ${\cal M}$ is given by:
$${\cal M}_{x,y}={1\over2}\sum_\mu\eta_\mu(x)
\left[\delta_{y,x+\hat\mu}-\delta_{y,x-\hat\mu}\right],\eqno(3.2)$$
where $\eta_\mu(x)$ are the Kawamoto-Smit phases $(-1)^{x_1+\cdots+x_{\mu-1}}$.

The Gross-Neveu model in two dimensions was first formulated using auxiliary
fields on the dual sites in reference [27]. We can motivate this particular
scheme by considering a unitary transformation to fields $u$ and $d$ [28]:
$$\eqalign{
u_i^{\alpha a}(Y)&={1\over{4\surd2}}\sum_A\Gamma_A^{\alpha a}\chi_i(A;Y), \cr
d_i^{\alpha a}(Y)&={1\over{4\surd2}}\sum_A B_A^{\alpha a}\chi_i(A;Y). \cr}
\eqno(3.3)$$
Here $Y$ denotes a site on a lattice of twice the spacing of the original, and
$A$ is a lattice vector with entries either 0 or 1, which
ranges over the corners of the elementary cube associated with $Y$, so that
each site on the original lattice corresponds to a unique choice of $A$ and
$Y$.
The $2\times2$ matrices $\Gamma_A$ and $B_A$ are defined by
$$\eqalign{
\Gamma_A&=\tau_1^{A_1}\tau_2^{A_2}\tau_3^{A_3},\cr
B_A&=(-\tau_1)^{A_1}(-\tau_2)^{A_2}(-\tau_3)^{A_3},\cr}\eqno(3.4)$$
where the $\tau_\mu$ are the Pauli matrices. Now, if we write
$$q_i^{\alpha a}(Y)=\left(\matrix{u_i^\alpha(Y)\cr d_i^\alpha(Y)\cr}\right)^a,
\eqno(3.5)$$
and interpret $q$ as a 4-spinor with two flavors counted by the latin index
$a$,
then the fermion kinetic term of the action (3.1) may be recast in Fourier
space as follows:
$$S_{kin}=\int{{d^3k}\over{(2\pi)^3}}\sum_i\sum_\mu{i\over2}\bigg\{
\bar q_i(k)(\gamma_\mu\otimes1_2)q_i(k) \sin2k_\mu+
\bar q_i(k)(\gamma_4\otimes\tau_\mu^*)q_i(k)(1-\cos2k_\mu)\bigg\},\eqno(3.6)$$
where
$$(\gamma_\mu)_{\alpha\beta}=
\left(\matrix{\tau_\mu&\cr&-\tau_\mu\cr}\right)_{\alpha\beta};\;\;
(\gamma_4)_{\alpha\beta}
=\left(\matrix{&-i1_2\cr i1_2&\cr}\right)_{\alpha\beta};\;\;
(\gamma_5)_{\alpha\beta}
=\left(\matrix{&1_2\cr 1_2&\cr}\right)_{\alpha\beta},\eqno(3.7)$$
the second set of ($2\times2$) matrices in the direct product act on the flavor
indices, and the momentum integral extends over the range
$k_\mu\in(-\pi/2,\pi/2]$. At non-zero temperature the lattice has finite
extent in the temporal direction, and $\int dk_0$ is replaced by a sum over
$N_\tau /2$ modes, where $N_\tau$  is the number of lattice spacings in the
time direction, and antiperiodic boundary conditions are imposed on the fermion
fields. In the classical continuum limit lattice spacing $a\to0$, the flavor
non-diagonal terms vanish as $O(a)$, and we recover the standard Euclidian
form $\bar q_j\partial{\!\!\! /}\,q_j$, where the flavor index
$j$ now runs from 1 to $N_f$.

The interaction term between the fermions and the auxiliary requires a little
care. We label the dual site $(x+{1\over2},x+{1\over2},x+{1\over2})$ by
$(A;\tilde Y)$ where $x$ corresponds to $(A;Y)$. In this particular labelling
the $\sigma$ fields are not all equivalent. For instance, for $A=(0,0,0)$ it
is easy to show the interaction term transforms to
$$\sigma(0;\tilde Y)\bar q_i(Y)(1_4\otimes1_2)q_i(Y).\eqno(3.8)$$
For $A=(1,0,0)$, however, it is more complicated:
$$\sigma(1;\tilde Y)\left[\bar q_i(Y)(1_4\otimes1_2)q_i(Y)
+{1\over2}\Delta^+_1\big (\bar q_i(Y)(1_4\otimes1_2+
i\gamma_1\gamma_4\otimes\tau_1^*)q_i(Y)\big  )\right],\eqno(3.9)$$
where $\Delta^+_\mu f(Y)\equiv f(Y+\hat\mu)-f(Y)\simeq2a\partial_\mu f(Y)$. We
see that there is an extra term containing non-covariant and flavor non-singlet
interactions, which is formally $O(a)$. If
we used a formulation in which the $\sigma$ fields lived on the direct lattice
sites, then such non-covariant terms would contribute at $O(a^0)$ [27].
It is straightforward to extend this analysis to all $A$; we conclude that the
interaction term may be written
$$S_{int}=\sum_Y\left(\sum_A\sigma(A;\tilde Y)\right)\bar q_i(Y)(1_4\otimes1_2)
q_i(Y)+O(a),\eqno(3.10)$$
which is clearly of the same form as the equivalent term in (1.2). In principle
the $O(a)$ non-covariant terms could be important once quantum loop corrections
are computed, and should be analyzed further. In the two-dimensional lattice
Gross-Neveu model their effect is discussed in [29], where it is argued that in
a perturbative expansion, even the on-site auxiliary formulation (with
non-covariances present at $O(a^0)$) yields the correct physics, that is with
no important continuum symmetries violated. This analysis appears to be readily
extended to the $1/N_f$ expansion in the three dimensional case.

Thus we see that the lattice action (3.1) reproduces the bosonised Gross-Neveu
model, at least in the classical continuum limit. Most
importantly, (3.1) has a discrete ${\rm Z}_2$ global invariance under
$$\chi_i(x)\mapsto(-1)^{x_1+x_2+x_3}\chi_i(x);\;\;
\bar\chi_i(x)\mapsto-(-1)^{x_1+x_2+x_3}\bar\chi_i(x);\;\;
\sigma({\tilde x})\mapsto-\sigma({\tilde x}),\eqno(3.11a)$$
ie.
$$q_i(Y)\mapsto(\gamma_5\otimes1)q_i(Y);\;\;
\bar q_i(Y)\mapsto-\bar q_i(Y)(\gamma_5\otimes1);\;\;
\sigma({\tilde x})\mapsto-\sigma({\tilde x}).\eqno(3.11b)$$
It is this symmetry, corresponding to the continuum form
$\psi\mapsto\gamma_5\psi$, $\bar\psi\mapsto-\bar\psi\gamma_5$, which is
spontaneously broken at strong coupling, signalled by the appearance of a
non-vanishing condensate $<\bar\chi\chi>$
or equivalently $<\bar q(1_4\otimes1_2)q>$. As discussed in Sec. 2, an
equivalent order parameter is the scalar field expectation $\Sigma_0$. To
leading order in $1/N_f$, we can compute the fermion tadpole explicitly to
yield the lattice gap equation, relating $\Sigma_0$ to $1/g^2$ (Cf. Eq.(2.4)).
Due to the simplicity of the loop integral, the contributions from the
$O(a)$ terms in (3.10) vanish, and we find
$${\Sigma_0\over g^2}={1\over V}{\rm tr}S_F=
\int_{\pi/2}^{\pi/2}{{d^3k}\over{(2\pi)^3}}{\rm tr}
{{-{i\over2}\sum_\mu[\sin2k_\mu(\gamma_\mu\otimes1_2)+
(1-\cos2k_\mu)(\gamma_4\otimes\tau_\mu^*)]+\Sigma_0(1_4\otimes1_2)}\over
{\sum_\mu\sin^2k_\mu+\Sigma_0^2}}.\eqno(3.12)$$
Using standard techniques, we arrive at a form suitable for numerical
quadrature:
$${1\over g^2}=\int_0^\infty d\alpha e^{-\alpha({\textstyle{3\over2}}+
\Sigma_0^2)}I_0^3({\alpha\over2}),\eqno(3.13)$$
where $I_0$ is the modified Bessel function. A plot of $\Sigma_0$ vs.
$1/g^2$ is shown in Fig. 5 -- we see that in this regularisation scheme, to
leading order in $1/N_f$, the bulk critical point $1/g^2_c\simeq1.011a^{-1}$.

The action (3.1) was numerically simulated using the hybrid Monte Carlo
algorithm [30], in which the Grassmann fields are replaced by real bosonic
pseudofermion fields $\phi(x)$ governed by the action
$$S=\sum_{x,y}\sum_{i,j=1}^{N_f/2}{1\over2}\phi_i(x)\big(M^tM\big)_{xyij}^{-1}
\phi_j(y) + {N_f\over{4g^2}}\sum_{\tilde x}\sigma^2({\tilde x}),\eqno(3.14)$$
where
$$M_{xyij}={\cal M}_{xy}\delta_{ij}+\delta_{xy}\delta_{ij}{1\over8}
\sum_{<{\tilde x},x>}\sigma({\tilde x}).\eqno(3.15)$$
Note $M$ is strictly real, and that in this model we are able to work directly
in the chiral limit bare fermion mass $m=0$, since the matrix $M$ has
non-vanishing diagonal entries, and can always be inverted.
Integration over $\phi$ yields the
functional measure $\sqrt{{\rm det}(M^tM)}\equiv{\rm det}M$ if the determinant
of $M$ is positive semi-definite. This condition is fulfilled if $N_f/2$ is an
even number.

The hybrid Monte Carlo algorithm works by evolving the fields through a
fictitious ``simulation time'' $\tau$ using a Hamiltonian
$H=\pi^2/2+S[\sigma]$,
where $\pi({\tilde x})$ is a momentum field conjugate to $\sigma$.
The fields are
sampled after evolving through a period of fixed or variable $\tau$ known as a
trajectory. Detailed balance is maintained by accepting or rejecting the entire
trajectory according to a Metropolis step calculated using the probability
weight $\exp(-\Delta H)$, and ergodicity by regenerating the $\{\pi\}$ after
each trajectory using gaussian random numbers. The Hamiltonian dynamics is
simulated by finite difference equations parameterized by some fundamental
interval $d\tau$: in the limit $d\tau\to0$ we expect the dynamics to conserve
energy, and hence the Metropolis acceptance rate to approach one. The art of
hybrid Monte Carlo lies in tuning the parameters of the simulation so that
$d\tau$ can be made as large as possible, to reduce the amount of computer time
per trajectory, whilst maintaining a reasonably high acceptance rate. One
method, as originally discussed in [30], is to alter the couplings of the
``guidance Hamiltonian'' used to evolve the system through the trajectory;
only the ``acceptance Hamiltonian'' used in the Metropolis step need be exact.
The idea is that couplings, masses, etc. may receive $O(d\tau)$ corrections,
so that the algorithm is in effect producing exact Hamiltonian dynamics at a
different point in coupling space.  In the work presented here we have
found that the acceptance rate of the algorithm may be optimized by tuning the
parameter $N_f$ in the action (3.14), so that the guidance Hamiltonian uses
$N_f^\prime>N_f$.
Interestingly, because quantum fluctuations are
$O(1/N_f)$, the amount of computer time needed to achieve comparable
statistical
accuracy is roughly independent of $N_f$: the only requirement that increases
with $N_f$ is the storage of the psuedofermion fields.

The hybrid Monte Carlo algorithm proved to run sufficiently quickly and
efficiently that unusually quantitative results could be obtained for large
enough $N_f$. Considerable details about the runs will be given below as
different computer experiments are described, but a few words about the
parameters used in the algorithm should be recorded here. In order to maintain
a high acceptance rate so the algorithm would produce configurations that
explored phase space rapidly, we tuned both $N_f^\prime$ and $d\tau$. Typically
as the lattice size was increased $d\tau$ had to be taken smaller and
$N_f^\prime$ approached $N_f$. For example, on a $6^3$ lattice when simulating
the $N_f=12$ theory, the choices $N_f^\prime=13.35$ and $d\tau=0.25$ gave
acceptance rates greater than 93\% for all couplings of interest. To maintain
this acceptance rate on a $12^3$ lattice we picked $N_f^\prime=12.38$ and
$d\tau=0.125$, and on a $24^3$ lattice $N_f^\prime=12.17$ and $d\tau=0.050$.
Measurements were taken only after a reasonable time interval had passed --
it is pointless to take measurements after each sweep because sequential
measurements are highly correlated, but measurements should not be taken so
infrequently that information is lost. For most of our runs we chose a
trajectory length $\tau$ equal to half a time unit -- in other words
$d\tau$ multiplied by the number of sweeps between refreshment was chosen to be
0.50. Note that on the $24^3$ lattice a trajectory rises to 10 sweeps and the
runs are proportionally more compute intensive than on small lattices. Of
course, we never assumed that sequential trajectories were statistically
independent. We used standard binning methods to calculate the variances in
our measurements for each observable. In a typical run of 5,000 trajectories,
say, a list of 5,000 measurements of the order parameter, etc. would be made
and
those lists would be analyzed to find the number of truly statistically
independent measurements. Near the critical point we found the usual symptoms
of critical slowing down -- tens of trajectories were needed to decorrelate
measurements of observables such as the order parameter. Variances could be
calculated after each run and the run extended if greater accuracy were needed.
In our tables of results to be discussed below, we typically list the number
of trajectories used and the statistical error bars for our raw data sets. Runs
as long as 10,000 trajectories were performed near the sytem's critical point
to extract scaling laws and critical indices with good accuracy and confidence.
Runs of this length are ten times as long as those currently practised in
lattice QCD. We have such good statistics here because of the lower
dimensionality of our model and the fact that each sweep has much fewer
floating point operations. In addition, the number of sweeps of the conjugate
gradient routines needed to effectively invert the lattice Dirac operator
during
each sweep is much smaller than in lattice QCD.
Typically, only 10 - 100 conjugate
gradient sweeps are needed for each inversion, while in lattice QCD ten times
as
many are needed in the present generation of simulations. The reason for the
difference presumably lies in the fact that Eq.(1.1) has only a discrete
rather than a continuous chiral symmetry. We very carefully monitored the
convergence of the various conjugate gradient routines we used. There was the
conjugate gradient used in the guidance Hamiltonian, that used in the
acceptance Hamiltonian, and that used in measurements of the chiral condensate.
The stopping residuals were typically chosen per lattice site to be
$10^{-4}$, $10^{-6}$, and $10^{-4}$ respectively. Since conjugate gradient
routines are only approximate, they could introduce an unwanted systematic
error into the algorithm. Therefore, we carefully checked that our observables
were insensitive to the size of the stopping residuals. Since the conjugate
gradient algorithm converges monotonically at an exponential rate, it is
relatively cheap (in computer time) to choose the stopping residual very
safely and conservatively. Lengthy test runs were made to assure ourselves
that all was in order.

We have not chosen to explore different values of $N_f$ systematically in this
study, but rather to thoroughly explore the system's critical behavior with the
choice $N_f=12$. However, in pilot studies on small lattices we did comparative
runs at $N_f=6$, 12, and 24. The results from a $12^3$ lattice for the
expectation value $\Sigma_0=<\sigma>$ vs, $1/g^2$ are plotted in Fig. 5. We
see that chiral symmetry is indeed spontaneously broken, and for $1/g^2$
between 0.5 and 0.8 the measurements are in fair agreement with the leading
order prediction (3.13): the points lie systematically below the line, which we
interpret as being due to $O(1/N_f)$ corrections as predicted by Eq.(2.19).
In Table II we give values for $\Sigma_0(N_f=\infty)$ given by (3.13) together
with the measured normalized deviations $N_f(\Sigma_0(\infty)-\Sigma_0(N_f))$.
The numbers for $N_f=24$ and 12 are consistent within errors, implying that the
deviation is $O(1/N_f)$. For $N_f=6$ the deviation is consistently larger,
possibly because for this small value $O(1/N_f^2)$ effects are becoming
significant. A similar trend was found in a high-statistics study of the
two-dimensional Gross-Neveu model [31]; here $N_f$ was varied between 4 and
120, and the $O(1/N_f)$ correction calculated exactly using a lattice
regularization. It was found that an $O(1/N_f^2)$ term was required to
accurately fit data for $N_f\leq12$. More work is needed in three dimensions
before reliable conclusions can be drawn; for one thing, at $N_f=6$ the model
does not have a positive definite determinant, and runs at adjacent values of
$N_f$ will be needed before the results of Table II can be taken seriously.

Finally we note that for $N_f=6,12$, chiral symmetry is restored for
$1/g^2\geq0.9$, and for $N_f=24$ it is restored for $1/g^2\geq1.0$. Finite
volume effects to be discussed in the next section make the precise
determination of the critical coupling $1/g_c^2$ difficult. In the rest of
the paper we will restrict our attention to the case $N_f=12$, and concentrate
on high precision studies in the region $1/g^2\geq0.7$, using a much finer
grid of $1/g^2$ values. This will enable a quantitative description of the
critical scaling properties of the model.
\vskip 1.5 truecm
\noindent{\bf4.
Symmetric Lattice Simulations and the Bulk Critical Point}

Our emphasis in this section is the discovery of the critical point
predicted by the large-$N_f$ expansion and a numerical calculation of its
critical indices, $\beta$ (magnetic), $\gamma$ (susceptibility) and $\nu$
(correlation
length).  This program will be very successful for $\beta$ and $\gamma$ while
an
accurate determination of $\nu$ will only occur once the behavior of the model
at non-zero fermion density has been investigated [32].
To find the critical point we can simply calculate $\Sigma_0$,
the vacuum
expectation value of the auxiliary field.  Since $\Sigma_0$ is proportional
to the chiral condensate $<\bar\psi\psi>$
and since its vacuum expectation value spontaneously breaks
chiral symmetry, it is a particularly convenient order parameter.  On small
lattices, however, vacuum tunneling processes which take $\Sigma_0$ to
$-\Sigma_0$ can
obscure the critical point.  The same problem affects computer simulation
calculations of the magnetization in the Ising model, for example, and can
be controlled by adding a small symmetry breaking field to the action.  We
chose not to do that here, although later we shall use this trick to
measure the critical index $\delta$.  Instead, we monitored each computer
simulation run for vacuum tunnelling events.  Away from the critical point
in the symmetry-broken phase such events were so rare that good
measurements of $\Sigma_0$ and its susceptibility $\chi$, given by the
variance of $\Sigma_0$, were possible.
There are presumably two reasons for this.
First, we chose $N_f=12$ which is sufficiently large that fluctuations
and tunnelling processes were unlikely.  And, second, our lattices were
sufficiently large ($8^3$ -- $20^3$) that the probability of tunnelling events
was highly suppressed.  However, in the immediate vicinity of the critical
point tunnelling events were sufficiently common that our computer data was
not useful.  Luckily we shall see that the scaling laws (critical indices)
we seek were apparent in the data over regions of coupling where tunnelling
was not a problem.

In Figs. 6 -- 9 we show the data for $\Sigma_0$ vs. $1/g^2$ and the reciprocal
of the susceptibility $1/\chi$ vs. $1/g^2$
on both the ordered and disordered sides of the
transition.  Our raw computer data, statistical errors and number of
trajectories of the hybrid Monte Carlo algorithm used for each measurement
can be found in Tables III -- VI.  For $1/g^2<1/g^2_c$, the critical point, we
expect a non-analytic vanishing of $\Sigma_0$,
$$\Sigma_0=C\left({1\over g^2_c}-{1\over g^2}\right)^\beta,\eqno(4.1)$$
where $\beta$ is the magnetic critical index.  This simple power behavior is
only expected for couplings $1/g^2$ sufficiently near the critical point
$1/g^2_c$. In leading order of the large $N_f$ expansion $\beta=1$.
Similarly the susceptibility should diverge at $1/g^2_c$;
$$\eqalign{
\chi&=A\left({1\over g^2_c}-{1\over g^2}\right)^{-\gamma}\;\;\;
g^2>g^2_c\cr
\chi&=B\left({1\over g^2}-{1\over g^2_c}\right)^{-\gamma}\;\;\;
g^2_c>g^2\cr}\eqno(4.2)$$
with the same index $\gamma$ above and below the transition.
In leading order of the large $N_f$ expansion $\gamma=1$.

The data plotted in Figs. 6 -- 9 are beautifully fitted by the large $N_f$
predictions $\beta=1$ and $\gamma=1$.  We are unable to extract the small
finite $N_f$ correction (for $N_f=12$,
$\gamma=1+8/N_f\pi^2+O(N_f^{-2})\simeq1.068$, $\beta=1+O(N_f^{-2})$)
expected in this simulation.  It may be necessary to work very
close to the critical point to find evidence for $\gamma=1.068$ rather than
$\gamma=1$, and that would require larger lattices to evade vacuum tunnelling
apparent in the figures and the Tables III -- VI which accompany them. Note
that
the error bars on the measured $\Sigma_0$ values are smaller than the circles
themselves of the figures.  The error bars on the susceptibility $\chi$
are considerably larger (5 -- 10\%) because it measures the width of the
fluctuations in the order parameter.  In Fig. 10 we plot $\ln\Sigma_0$
vs. $\ln(1/g^2_{c20}-1/g^2)$
where $1/g^2_{c20}\simeq0.955$ is the best estimate for the critical coupling
of a $20^3$ lattice inferred from Fig. 9.  The straight-line fit in Fig. 10
gives $\beta=1.00$ and is in almost perfect agreement with the measurements at
couplings $1/g^2=0.70$ -- 0.825. Comparing Tables V and VI we see that over
these
couplings the $\Sigma_0$ measurements are in agreement to better than 1\% on
the $16^3$ and $20^3$ lattices.
Closer to the critical point we do not have comparable
control over finite size effects.  It would be interesting to redo Fig. 10
on a larger lattice and see if the $\Sigma_0$ values at
$1/g^2>0.825$ approach the
scaling curve.  Similar curves for the susceptibility $\chi$ can be made
($\ln\chi$ vs. $\ln(1/g^2_{c20}-1/g^2)$)
and the critical index $\gamma$ found to be 1.0(1).  Much
greater statistics would be needed to reduce the error on the determination
of $\gamma$ to one percent where the finite $N_f$ corrections in Eq. (1.3)
could be probed. A major barrier to our obtaining a higher precision
determination of $\gamma$
is the uncertainty in the exact critical coupling and the fact that our
estimates of it are hindered by vacuum tunnelling.

One notices from Figs. 6 -- 9 that the peak of the system's susceptibility
shifts with the linear lattice size $L$.  Finite size scaling arguments
relate the size dependence to the correlation length exponent $\nu$;
$${1\over{g^2_c(L)}}-{1\over{g^2_c}}=aL^{-{\textstyle{1\over\nu}}},\eqno(4.3)$$
where $1/g^2_c(L)$ is the coupling where $\chi$ peaks on a $L^3$ lattice
and $1/g^2_c$ is the
infinite volume limit.  Unfortunately the results shown in Figs. 6 -- 9 do not
determine $1/g^2_c(L)$ accurately enough to determine $\nu$ well.
In Table VII we give the $1/g^2_c(L)$
values with error bars and in Fig. 11 we plot Eq. (4.3) for $\nu=1$
and infer $1/g^2_c\simeq1.00$.  In the next section of this article
we shall study the four-fermi model on asymmetric
lattices (nonzero temperature) and will achieve quantitative results for $\nu$.
\vskip 1.5 truecm

\noindent{\bf5. Asymmetric Lattice Simulations, the Critical Temperature,
and the Index $\nu$}

We simulated the four-fermi model on asymmetric lattices,
$N_\tau\times N^2$, to
determine its critical termperature $T_c$ where chiral symmetry is restored
and to determine the order of this transition.  Lengthy runs, several tens
of thousands of trajectories of the hybrid Monte Carlo algorithm, in the
vicinity of the phase transition on lattices with $N_\tau$ ranging from 2 to 12
and $N$ set to $3N_\tau$ showed no convincing evidence for metastability or
tunnelling between a symmetric and an asymmetric vacuum.  So, we concluded
that the transition was second order in agreement with the large $N_f$
analysis of this transition [17,32].  In Fig. 12 we show the histograms of
measurements of $\Sigma$ (we reserve the subscript zero for the zero
temperature value) in the ground state at couplings $1/g^2=0.86$, 0.865 and
0.870 on a $10\times30^2$ lattice.  At $1/g^2=0.860$ the system is in a
chirally broken
phase with $\Sigma$ very small, 0.039(1), but definitely nonzero.
At $1/g^2=0.865$
vacuum tunnelling occurs between two states with $\Sigma=\pm0.030$, and
finally, at $1/g^2=0.870$ the distribution of $\Sigma$ measurements is centered
around the origin indicating symmetry restoration. For $1/g^2$ values
decreasing
below $1/g^2_{\beta c}(N_\tau=10)=0.865(5)$, the mean values of
$\Sigma$ increase smoothly, which is
indicative of a second order phase transition.  Since $T=1/N_\tau=0.10$ in
lattice units, we see that the critical temperature measured in units of
$\Sigma_0$ is $T_c/\Sigma_0=0.64(4)$ in this case.  (Note from Fig. 9 that
$\Sigma_0=0.157(1)$ at coupling $1/g^2=0.865$ at zero temperature.)
For values of $1/g^2$
greater than $1/g^2_{\beta c}(N_\tau=10)$ the mean values of
$\Sigma$ on the $10\times30^2$
lattice were always consistent with zero and histograms of particular runs
were well fitted with normal distributions centered at zero.  Similar
investigations on lattices with $N_\tau$  ranging from 2 to 12 led to the
estimates of the critical couplings $1/g^2_{\beta c}(N_\tau)$ recorded in Table
VIII. Using Fig. 9 to read off zero temperature vacuum expectation values of
$\Sigma$ at these couplings produces
the estimates of $T_c/\Sigma_0$ shown in Table IX and plotted in Fig. 13.

These results should be compared to the large $N_f$ prediction [17,32],
$${T_c\over\Sigma_0}={1\over{2\ln2}}\simeq0.72,\eqno(5.1)$$
which is plotted in Fig. 13.  Our $N_\tau=12$ result is within a standard
deviation of the exact result and the trend of the simulation results,
$T_c/\Sigma_0$ increasing slowly with $N_\tau$, is quite satisfactory.

Next we can use the data in Table VIII to obtain a better estimate of the
correlation length index $\nu$.  The shifts of the critical couplings
$1/g^2_{\beta c}(N_\tau)$ with lattice extent in the temporal direction should
scale as
$${1\over{g^2_{\beta c}(N_\tau)}}-{1\over g^2_c}=
bN_\tau^{-{\textstyle{1\over\nu}}},
\eqno(5.2)$$
where $1/g^2_c$ is the large volume limit of the bulk critical coupling. From
Fig. 10 we know that $1/g^2_c$ is between 0.95 and 1.00, but unfortunately
cannot be specified with any greater precision.
In Fig. 14 we plot Eq. (5.2) with the data of Table VIII
for the choices of $1/g^2_c=0.950$, 0.976 and 0.995.  In each case linear fits
of $\ln(1/g^2_c-1/g^2_{\beta c}(N_\tau))$ vs. $\ln N_\tau$ are possible for
$N_\tau=6$, 8, 10 and 12.  The
slopes determine the critical index $\nu$, and we have from the fits
$\nu=0.81$ if $1/g^2_c=0.950$,
$\nu=0.94$ if $1/g^2_c=0.976$ and $\nu=1.07$ if $1/g^2_c=0.995$.  Clearly our
precision for a good determination of $\nu$ is limited by our uncertainty in
the critical coupling.  Nonetheless, our best estimate for $\nu$,
$$\nu=0.94(13),\eqno(5.3)$$
is consistent with and close to the large $N_\tau$ prediction of unity.  In the
second paper of this series [32] where we study the four-fermi model at nonzero
chemical potential a more accurate and independent determination of $\nu$ will
be given which lies within the error bars of Eq. (5.3).
\vskip 1.5 truecm

\noindent{\bf6.
The Critical Index $\delta$ for the Bulk and the Finite Temperature Transition}

The critical index $\delta$ controls the response of a magnetic system's
magnetization $M$ at criticality to a small external magnetic field $h$:
$$M\vert_{T=T_c}\propto h^{\textstyle{1\over\delta}}.\eqno(6.1)$$
For the four-fermi model $M$ maps into the chiral order parameter $\Sigma_0$
(or $<\bar\psi\psi>$) and $h$ maps into a bare fermion mass (which explicitly
breaks the chiral symmetry), so
$$\Sigma_0\vert_{g^2=g^2_c}\propto m^{\textstyle{1\over\delta}}.\eqno(6.2)$$
For large $N_f$ the index $\delta$ is predicted to be 2, Eq.(1.3).
We attempted
to measure $\delta$ by introducing a bare fermion mass into the theory's
action,
and simulating the system on $8^3$, $16^3$ and $24^3$ lattices at criticality
but variable $m$.  Since we do not know $1/g^2_c$ exactly we did these
simulations at $1/g^2=1.00$ and
0.975.  In addition, since we do not know beforehand how small $m$
must be taken to see the scaling law Eq. (6.2), we simulated $m$ values
ranging from 0.00375 to 0.25 in lattice units and searched numerically for a
simple scaling form, Eq. (6.2).  We found that very accurate measurements
of $\Sigma_0$ were possible.  Vacuum tunnelling was always suppressed in these
measurements because of the explicit symmetry breaking in the action.  Our
computer results are given in Tables X and XI.

In Fig. 15 we
plot the data for a sequence of lattices $8^3$, $16^3$ and $24^3$.  In
fact, $\ln\Sigma_0$ is plotted against $\ln m$ for $1/g^2=1.00$. We see
finite size effects at the smallest $m$ values as expected -- as $m$ decreases
larger and larger lattices are needed to measure $\Sigma_0$'s nonzero value.
For $m$ ranging from 0.0625 to 0.01625 we see the $8^3$ and $16^3$ data
approaching the $24^3$ data from above.  A linear fit over this range of $m$
gives $\delta=2.0$ in almost perfect agreement with the large $N_f$
predictions.
Note that fermion masses less than 0.01625 were not used in the fit because
the $16^3$ and $24^3$ lattices do not give consistent values for $\Sigma_0$,
indicating
that finite size effects are still important here.  Presumably, if one
simulated $32^3$ lattices, a wider ``scaling window'' would be seen.  Note that
error bars on the $\Sigma_0$ measurements are smaller than the symbols in the
figure.

In Fig. 16 we show identical plots for simulation runs at $1/g^2=0.975$ which
could also be the bulk critical point.  The value obtained for $\delta$ for the
fit here is slightly larger, 2.21, than that discussed above.  We again see
that without better knowledge of the bulk critical coupling our uncertainty
in the critical index $\delta$ will be about 10\%. Uncertainty in the position
of critical points usually limits the precision of finite size scaling studies
in other statistical mechanics models as well.

Finally, we measured $\delta$ for the finite temperature rather than the bulk
critical point.  This is particularly interesting because the large $N_f$
prediction for $\delta$ associated with the finite temperature transition is 3
[17],
indicative of a Gaussian model rather than the nontrivial scaling theory at
the bulk critical point.  To determine this critical index we used the
largest lattice studied in Sec. 4, $12\times36^2$ and added a bare mass term to
the action.  In this case the critical coupling for the finite temperature
transition on the $12\times36^2$ lattice was determined accurately in Sec. 4,
$1/g^2_{\beta c}(N_\tau=12)=0.880(5)$,
so a relatively high precision measurement of its $\delta$ is
possible.  The computer data are recorded in Table XII. In Fig. 17 we show
$\ln\Sigma_0$ vs. $\ln m$ for $m$ values ranging from 0.01 to 0.000625.
Linear fits to
the data produce the result $\delta=3.0(1)$ in excellent agreement with our
analytic expectations.
\vskip 1.5 truecm
\noindent{\bf7. The Scalar Correlation Function}

In this section we report on measurements of the scalar two-point function
$<\sigma(0)\sigma(x)>$, and our attempts to understand them using the
predictions
of the $1/N_f$ expansion described in Sec. 2.  Since we have been interested
for the most part in the bulk properties of the model, we did not perform any
special simulations on lattices which are longer in one direction, which is
normal practice in numerical QCD, where the requirement is to examine the
asymptotic behavior of hadronic correlators in order to extract hadron masses.
Further studies may benefit from this approach: however, as we shall see, the
relation between the asymptotic decay of the correlator and the particle mass
is by no means so straightforward in this model, and indeed, in the
symmetric phase we are only able to extract a width by fitting to the
sub-asymptotic behavior! Instead, we performed
measurements on $20^3$ lattices at the values of the coupling $1/g^2$ used
elsewhere.

As usual, in order to simplify the fit and maximize statistics we projected
on to the zero spatial momentum sector by summing over timeslices, and define
the correlator:
$$C(x)={1\over N_\tau}\sum_{\tau=1}^{N_\tau}{1\over V_s^2}
\left<\left(\sum_{\vec y}\sigma(\vec y;\tau)\right)\left(\sum_{\vec y}
\sigma(\vec y;\tau+x)\right)\right>,\eqno(7.1)$$
where $\vec y$ labels the $V_s$ sites existing on each of $N_\tau$ timeslices.
Of course, the quantity of interest is really the connected correlator
$$C_c(x)=C(x)-{1\over V_s^2}<\sum_{\vec y}\sigma(\vec y;0)><\sum_{\vec y}
\sigma(\vec y;x)>.\eqno(7.2)$$
In our work we assume the expectation value $<\sigma>$ is translationally
invariant, and extract $<\sigma>^2$ as a fitted parameter in fits of $C(x)$.
This is a much more stable procedure than subtracting the measured value
$\Sigma_0^2$ from the data before fitting, because the vacuum and the other
quantities entering the fit                are highly correlated. Of course, we
expect agreement between the two quantities.

In order to parameterize the decay of $C_c(x)$, we use the leading order
expressions for $D_\sigma$ given in Sec. 2. First we discuss fits in the
broken phase, where the scalar is expected to be a genuine massive bound state.
{}From (2.7) we have
$$C_c(x)\propto\int_{-\infty}^\infty dk{{e^{ikx}}\over{(k^2+M^2)
F(1,{1\over2};{3\over2};-{k^2\over M^2})}},\eqno(7.3)$$
where here $M$ denotes the scalar mass.
Immediately we see the difference between this
model and QCD: the presence of the hypergeometric function in the denominator
introduces branch cuts in the complex $k$-plane along the imaginary $k$-axis
in the ranges $[iM,i\infty)$ and $(-i\infty,-iM]$. Instead of a discrete series
of poles, corresponding to a set of bound states in the channel in question,
which
is the case in a confining theory such as QCD, we have a complicated spatial
dependence due to a strongly-interacting  fermion -- anti-fermion continuum
for $k>M_\sigma=2\Sigma_0$ (to leading order in $1/N_f$) [9].
Performing the contour
integral around the branch cut in the upper half-plane, we find
$$\eqalign{
C_c(x)&\propto P(x;M)=\int_1^\infty dv{{ve^{-Mvx}}\over{(v^2-1)[\pi^2+
4(\coth^{-1}v)^2]}}\cr
&=\int_0^1dt{{(1+e^{-1/t})}\over{t^2(2+e^{-1/t})}}{{\exp(-Mx(1+e^{-1/t}))}
\over{[\pi^2+\ln^2(1+2e^{1/t})]}}+
\int_0^{e/e+1}du{1\over{u(1-u^2)}}{{e^{-Mx/u}}\over{[\pi^2+4(\tanh^{-1}u)^2]}},
\cr}\eqno(7.4)$$
where the second form is suitable for numerical quadrature. Due to the
contribution of the fermion -- anti-fermion continuum, $P(x;M)$ decays faster
than $e^{-Mx}$, and the corrections to the latter form cannot be expressed as
a discrete sum of exponentials over higher states. This means that we no
longer have methods
such as a search for a plateau in the effective mass plot to tell us when the
form (7.4) might become valid: we must simply trust its  validity over the
whole range of $x$, or at least that range exceeding the correlation length,
where we might expect the continuum approximation to be accurate. Note that in
ref. [31] the scalar correlator in the two-dimensional Gross-Neveu model,
which is asymptotically free, was analyzed in this way, and it was found that
the inclusion of a free fermion continuum in the fit did improve the results.
In our case, however, the fermions are strongly interacting, so we are forced
to try (7.4).

We have fitted $C(x)$ for $1/g^2$ ranging from 0.7 to 0.975, using the
three-parameter form $a[P(x;M)+P(L-x;M)]+\Sigma_0^2$, where $L=20$ is the
length of the lattice. The results are presented in Table XIII and plotted in
Fig. 18 (error bars are not shown for clarity). We used a least squares fit
with a correlated $\chi^2$ function [33] to allow for correlation between
timeslices. Errors were estimated using standard binning procedures. For
comparison we also tried a standard single pole fit of the form
$a[e^{-Mx}+e^{-M(L-x)}]+\Sigma_0^2$. In both cases we fitted for $x$ ranging
between 2 and 9; this produced $\chi^2$ values per degree of freedom $\leq1$ in
most cases (the maximum $\chi^2$ per d.o.f. found was 1.64).

We can make the following observations. The values of $M$ obtained by the
branch cut form $P(x;M)$ lie systematically below those obtained from the
single pole form, as expected: however the fluctuations in the two are clearly
correlated. We see that $M$ decreases as $g^2\to g^2_{c+}\simeq0.96$ on the
$20^3$ lattice, but does not reach zero as required by the hypothesis that it
represents an inverse correlation length. The points at $1/g^2=0.95$, 0.975
clearly lie outside this trend, probably because they are in or very close to
the symmetric phase. The scatter of the $M$ values about the downward trend is
accounted for by the errors given in Table XIII. We also plot the fitted values
of $2\Sigma_0$ (which coincide within error bars for both fits), which at
leading order in $1/N_f$ corresponds to the two fermion threshold (note that
even for a correlation length of 10, the $O(1/N_f)$ corrections predicted by
Eq.(2.15) amount to just 5\% for $N_f=12$).  The line drawn through these
points is that obtained from a fit to the data of Table VI, so we see the
fitted
values of $\Sigma_0$ coincide with good accuracy with the bulk averages, which
is a consistency check of our fitting method.

If the scalar field interpolates a true bound state, then $M\leq2\Sigma_0$. We
expect the difference, representing the binding energy, to be $O(1/N_f)$. We
see that our fits for $M$ using $P(x;M)$ obey  this criterion only for
$1/g^2<0.875$; moreover there is no clear region of the plot where the ratio
$M:2\Sigma_0$ is constant, so no estimate of $M$ in physical units can be made.
Given the systematic uncertainties in the fitting procedure highlighted by the
difference between the two fitted forms, we interpret this as an indication
that the leading order form (7.3,4) is still not adequate to fit the measured
correlator. Two further avenues suggest themselves. First, one could calculate
the leading order scalar propagator in an explicit lattice regularization:
this would probably need to be done numerically, but might, for instance,
modify  parameters such as the $\pi^2$ in the denominator of (7.4) which at
present are put into the fits ``by hand''. Secondly, one could bite the bullet
and attempt to calculate $D_\sigma(k^2)$ to $O(1/N_f)$ in the broken phase, in
order to be able to parameterize the separation between the bound state pole
and the two-fermion threshold. Our attempts to do this in a naive way did not
yield useful results.

Next we consider the scalar correlator in the symmetric phase. As discussed in
Sec. 2, in this case the scalar field does not interpolate a stable massive
particle state, but rather an unstable resonance. The resonance width $\mu$
serves as an inverse correlation length, and can be extracted from the data as
follows. We start from the definition (7.1) of $C(x)$ and the leading order
form for $D_\sigma(k^2)$ given in (2.26):
$$C(x)\propto\int_0^\infty{{\cos kx}\over{k+\mu}}={\rm Re}\left(e^{-i\mu x}
E_1(-i\mu x)\right)\equiv g(\mu x).\eqno(7.5)$$
Here $E_1(z)$ is the exponential integral function defined by [34]:
$$E_1(z)=\int_z^\infty{{e^{-t}}\over t} dt.\eqno(7.6)$$
For $\mu x \to\infty$ we have
$$g(\mu x)=     {1\over{(\mu x)^2}}\left(1+O({1\over{(\mu x)^2}})\right).
\eqno(7.7)$$
In this phase the scalar correlator decays asymptotically with a power law:
the corrections to this power law at small $\mu x$ enable us to extract the
physical scale $\mu$. Because of the power-law form, it is not sufficient to
sum over just forward and backward propagating signals to get an accurate fit:
one must sum over the signals from all ``images'' corresponding to propagation
an arbitrary number of times around the periodic lattice. Using the asymptotic
form (7.7) and the Euler-McLaurin formula, we can approximate this sum:
$$\eqalign{
C(x) &\propto Q(x;\mu)=\sum_{n=-\infty}^\infty g(\mu(x+nL))\cr
&=\sum_{n=-N}^N g(\mu(x+nL))+{1\over{\mu^2L}}\left({1\over{NL+x}}+{1\over{NL-x}
}\right)\cr&\phantom{\sum_{n=-N}^N g(\mu}
-{1\over{2\mu^2}}\left({1\over{(NL+x)^2}}+{1\over{(NL-x)^2}}\right)
+{L\over{6\mu^2}}\left({1\over{(NL+x)^3}}+{1\over{(NL-x)^3}}\right)+O(N^{-4}).
\cr}\eqno(7.8)$$

In Table XIV and Fig. 18 we show the results of the two-parameter fit
$C(x)=aQ(x;\mu)$ to results obtained for $1/g^2$ between 1.1 and 0.925. Once
again, we fitted to $x$ values 2 -- 9, and the $\chi^2$ per d.o.f. was in most
cases less than one, with a maximum value 1.58. Fits of the standard simple
pole form in this region were of worse quality and yielded higher values of
$\mu$: moreover if we constrained the vacuum $\Sigma_0=0$, then these fits
became disasterously poor. In most cases we found the series truncation in
(7.8)
was adequate for $N=5$. Only for the very smallest $\mu$ at $1/g^2=0.925$ did
we
find some weak dependence of the fit on $N$ -- here we quote the $N=40$ result.

{}From Fig. 18 we see that in the symmetric phase the fitted parameter $\mu$
decreases fairly smoothly to zero as $g^2\to g^2_{c-}$, and has every
appearance of a genuine inverse correlation length. The data are not yet of
sufficient quality to make a meaningful extraction of the exponent $\nu$,
however. It is also not clear how reliable the points at $1/g^2=0.925$ and
0.95 are -- perhaps a study on different sized lattices might be useful here.
Overall, however, given the limitations of our statistics and lattice sizes, we
 find the results of the fit quite satisfactory. This is probably a sign that
the leading order form of the propagator in the symmetric phase (2.26) is not
grossly changed by $1/N_f$ corrections: this is of course verified by (2.35),
where we find the only change to $D_\sigma(k^2)$ is an $O(1/N_f)$
correction to the power of $k$.

To sum up, we have identified a correlation length associated with the
auxiliary field $\sigma$ in both phases of the model, which increases as
$1/g^2\to1/g^2_{c\pm}$. Leading order $1/N_f$ calculations predict that
the inverse correlation length should decrease linearly in this limit in
either phase, ie. $\nu=1+O(1/N_f)$. In the symmetric phase our accuracy
appears limited
principally by statistics and lattice volume; the leading order
prediction is qualitatively verified, but we are unable to fit directly
for the exponent $\nu$. In the broken phase, the correlation length we
extract from our fits does not diverge, and we suspect an improved
 understanding of the $1/N_f$ or $O(a)$ corrections in this phase will be
required before more progress can be made.
\vskip 1.5 truecm
\noindent{\bf8. Summary}

In this article we have attempted to reconcile two alternative views of the
strongly-coupled critical point which is generic to the model four-fermi
Lagrangian (1.1). In the language of statistical mechanics, we have
characterized the fixed point by calculating its critical exponents, which in
turn define its universality class. The indices satisfy consistency conditions
known as hyperscaling relations, which supports the idea of a single divergent
correlation length in the critical region. The critical correlation functions
display non-canonical scaling, that is, the operators corresponding to
physical observables acquire anomalous dimensions.

In particle physics language, we have seen that for some sufficiently large
coupling, the most stable vacuum is one in which chiral symmetry is
spontaneously broken, leading to a dynamically-generated fermion mass. An
expansion in powers of $1/N_f$ about the critical point is exactly
renormalizable for a continuum of dimensionality $2<d<4$ -- this behavior is
unlike standard weak coupling perturbation expansions (WCPE) about the Gaussian
fixed point, which are exactly renormalizable only for a single critical value
of $d$. Interactions in the model occur via exchange of scalar degrees of
freedom. The composite nature of these particles ensures that the physical
interaction strength remains non-vanishing as the UV cutoff is sent to infinity
-- the theory is non-trivial.

We have argued, in general terms in section 1 and in more detail in section 2,
that these descriptions are equivalent, and that renormalizability,
hyperscaling
and non-triviality are inextricably linked in this model. The composite nature
of the $\sigma$ field is signalled by its large anomalous dimension. The main
results of the explicit calculations of $O(1/N_f)$ corrections in section 2
are as follows: firstly, at this order and beyond, logarithmic divergences
appear, which necessitate vertex, wavefunction, and mass renormalizations.
The set of primitively divergent diagrams is such that the renormalization
constants are overdetermined, which means that renormalizability depends on a
non-trivial cancellation between different one-particle irreducible (1PI)
vertex functions. This cancellation can be viewed as a constraint on the
critical indices equivalent to hyperscaling, but ultimately results from the
existence of Ward identities arising from the chiral symmetry of the model.
Secondly, we have seen that the critical exponents get explicit $1/N_f$
corrections, implying that theories with different $N_f$ belong to different
universality classes. As we calculate to higher order in $1/N_f$, we
improve our understanding of the fixed point theory -- this is in contrast to
WCPE, where the fixed point is known to be Gaussian from the start, and
perturbative corrections serve to parameterize trajectories in or near the
critical hypersurface, that is, to relate massless theories at differing
momentum scales. In the latter case we say the WCPE interaction is marginal,
whereas in the four-fermi model the interaction $(\bar\psi\psi)^2$ is relevant,
since any departure from the critical coupling strength $g^2\not=g^2_c$ results
in the theory becoming massive, with a finite correlation length, either in the
conventional sense ($t>0$), or by generating a dimensionful width for the
$\sigma$ resonance ($t<0$). So far we have no reason to suspect the critical
hypersurface has a dimension greater than zero [22]. In simple terms, the
$1/N_f$ expansion can never yield a ``running coupling constant'', since
$N_f$ is a basic parameter of the theory and is not itself renormalized.
Thirdly, as a result of the consistency between the different 1PI amplitudes,
or if preferred the chiral symmetry of the model, we see that in the deep
Euclidean limit $k^2\to\infty$ the four-fermi scattering amplitude has no
dependence on $1/N_f$, apart from a trivial overall factor,
and assumes a universal form $A_d/N_fk^{d-2}$. This is
the universal interaction which characterizes the UV fixed point. It is
interesting that despite the fact that we have used the $1/N_f$ expansion to
calculate radiative corrections, the $1/N_f$ corrections manifest themselves
in the IR properties of the theory, such as the bulk critical exponents,
and presumably the bound state spectrum, about which we have little knowledge
at present.
Finally, we have shown that the picture we have developed is readily extended
to models with a continuous chiral symmetry, so that there appear to be no
problems incorporating massless Goldstone excitations.

On the numerical side, our main result is that the leading order predictions of
the $1/N_f$ expansion have been verified in a non-perturbative simulation,
suggesting that the $1/N_f$ expansion is accurate at least for $N_f\geq12$. The
critical exponents $\beta$, $\gamma$, $\delta$ and $\nu$ have all been directly
estimated and found to be distinctly non-Gaussian -- this is strong evidence
for the presence of anomalous dimensions. Moreover, simulations at non-zero
temperature support a crossover from the leading order result $\delta=d-1$ to
the Gaussian mean field value $\delta=3$.
We have not been able to measure explicit
$1/N_f$ corrections in this study, but are able to estimate the extra
computational effort required. We have also laid the groundwork for simulations
at smaller $N_f$, where quantum fluctuations will be larger, and the $1/N_f$
corrections easier to detect. It is, of course, possible that for some
sufficiently small $N_f$ the picture we have developed breaks down and the
$1/N_f$ expansion loses its applicability -- one suggestion has been that the
transition becomes first order [35]. Another direction, which we have explored
in a separate paper [32], is to simulate the model at non-vanishing chemical
potential. Unlike the case for lattice gauge theories, the action (3.1)
remains real under this modification, which means that we are able to study
a relativistic theory of interacting fermions at non-zero density on far larger
systems than are otherwise possible. As reported in [32], these simulations
have enabled an independent estimate of the exponent $\nu$.

Progress on understanding the scalar two-point function has been slower,
highlighting the difficulties in understanding the spectrum of a theory which
is both strongly coupled and non-confining. The main novelty and success here
was in the symmetric phase, where a ``mass'' (inverse
correlation length) was extracted from a fit to the pre-asymptotic behavior of
the correlator. Further progress in this direction probably depends upon a more
complete understanding of $O(1/N_f)$ corrections in the broken phase, which is
technically much tougher than the calculation presented here. Of course, in the
long run one would wish for a complete understanding of the bound-state
spectrum.

We hope that many of the aspects we have learned about this model will be of
relevance to strongly coupled quantum electrodynamics in four dimensions,
which has been the focus of much recent work by ourselves and others. Both
models exhibit a continuous phase transition from a ``massless'' phase to one
in which mass is dynamically generated -- and both models appear to scale with
non-Gaussian critical exponents [36]. The great hope is, of course, that
$QED_4$
will turn out to be the first known example of a non-trivial non-asymptotically
free theory in four dimensions. As yet, we have only analytic arguments
based on
self-consistent approximations, and numerical simulations, which, however
powerful, are always in need of some kind of interpretation. In the models
discussed here we have the additional benefit of a systematic approximation,
which has enabled us to make precise calculations and gain physical insight.
Perhaps the ultimate result of this work will be the development of a concise
vocabulary to describe strongly-coupled UV fixed points.
\vskip 1.5 truecm
\noindent{\bf Appendix}

Here we give some technical details of how to perform various momentum
integrals
in $d$-space. First, we define our Dirac matrices for all values of $d$ to
be $4\times4$, traceless, and hermitian, so that the trace always yields a
factor of 4. Alternative definitions are possible [21], but the results are
qualitatively unchanged, at least to $O(1/N_f)$.

We define the integral over Euclidean $d$-momentum $p$ as follows:
$$\eqalign{\int_p f(p^2)=&{2\over{(4\pi)^{d\over2}\Gamma({d\over2})}}
\int_0^\Lambda p^{d-1}f(p^2) dp;\cr
\int_p p_\mu f(p^2)=0&;\;\;\;\int_p p_\mu p_\nu f(p^2)=
{{\delta_{\mu\nu}}\over d}\int_p p^2 f(p^2).\cr}\eqno(A.1)$$
Although the formul\ae\ are superficially similar, this is {\it not\/}
dimensional regularisation; the cutoff $\Lambda$ appears explicitly in (A.1)
and is used to regulate divergent integrals.
The following formula, which holds for $d\in(2,4)$, serves to evaluate loop
integrals involving only fermi propagators:
$$\int_p{1\over{(p^2+M^2)^n}}=\delta_{n1}{{2\Lambda^{d-2}}\over
{(4\pi)^{d\over2}(d-2)\Gamma({d\over2})}}+{{M^{d-2n}\Gamma(n-{d\over2})}\over
{\Gamma(n)(4\pi)^{d\over2}}}.\eqno(A.2)$$

Finally, we give a brief outline of how to evaluate the two-loop integral
(2.29). With the momentum routing specified the integral over $p$ is fairly
straightforward: after taking the trace it reduces to several components which
can each be brought to the form (A.2) using a suitable Feynman
parameterization and momentum shift. If only one Feynman parameter is required
in a particular sub-integral, it is then simple to extract divergent terms of
the power-law form (2.30).  If there are two or more Feynman parameters, then
there will be an intermediate integral with a denominator, which is a function
of $q$ and external momentum, raised to a non-integer power $s$. The
following formula is needed:
$$\eqalign{\int{{A+Bx+Cx^2}\over{(a+bx+cx^2)^s}}dx&=\left(A-{{bB}\over{2c}}+
{{C(b^2-2ac)}\over{2c^2}}\right)\left({{4c}\over\Delta}\right)^s
\left(x+{b\over{2c}}\right)F\left(s,{\textstyle{1\over2}};{\textstyle{3\over2}};
-{{4c^2}\over\Delta}\left(x+{\textstyle{b\over{2c}}}\right)^2\right)\cr
&+\left(B-{{bC}\over c}\right){{(4c)^{s-1}}\over{2(1-s)c\left[\Delta+4c^2\left(
x+{b\over{2c}}\right)^2\right]^{s-1}}}\cr
&+{{C\Delta}\over{4c^2}}\left({{4c}\over\Delta}\right)^s\left(x+{b\over{2c}}
\right)F\left(s-1,{\textstyle{1\over2}};{\textstyle{3\over2}};
-{{4c^2}\over\Delta}\left(x+{\textstyle{b\over{2c}}}
\right)^2\right),\cr}\eqno(A.3)$$
where $\Delta\equiv4ac-b^2$ and $F$ is the hypergeometric function. This result
is exact. However, it turns out that the momentum-dependent logarithmic
divergences (2.31)  naturally arise from the {\it lower\/} limit of the Feynman
parameter integration, ie. from setting $x=0$ above. Also, in practice, the
coefficients $b$, $c$ are $O(q^2)$, where $q$ is the remaining loop momentum,
but $a$ is $O(q^0)$. These two considerations make the following
simplifications
possible:
$$\eqalignno{&\int_0{{dx}\over{(a+bx+cx^2)^s}}=-{b\over{\Delta(s-1)a^{s-1}}}+
O\left({1\over q^4}\right);&(A.4a)\cr
&\int_0{{xdx}\over{(a+bx+cx^2)^s}}={2\over{\Delta(s-1)a^{s-2}}}+
{{(3-2s)}\over{\Delta(s-2)(s-1)a^{s-2}}}+O\left({1\over q^6}\right);&(A.4b)\cr
&\int_0{{x^2dx}\over{(a+bx+cx^2)^s}}=
{{2(3-2s)b}\over{\Delta^2(s-2)(s-1)a^{s-3}}}+{{2(5-2s)b}\over
{\Delta^2(s-3)(s-2)a^{s-3}}}+O\left({1\over q^8}\right).&(A.4c)\cr}$$
With judicious choice of Feynman parameterization these formul\ae\ suffice to
evaluate (2.29). For example, suppose after taking the trace, introducing
Feynman parameters $x$ and $y$, and shifting the momentum $p$, we are left with
the following  $p$ sub-integral:
$$I(q,k)=8\int_0^1dx\int_0^{1-x}dy\int_p{{xk^2+yk.q}\over
{[p^2+x(1-x)k^2+y(1-y)q^2-2xyk.q]^3}}.\eqno(A.5)$$
We can do the integration over $p$ using (A.2):
$$I(q,k)=4{{\Gamma(3-{d\over2})}\over{(4\pi)^{d\over2}}}\int_0^1dx\int_0^{1-x}dy
{{xk^2+yk.q}\over{[x(1-x)k^2+(q^2-2xk.q)y-q^2y^2]^{3-{d\over2}}}}.\eqno(A.6)$$
Now use (A.4) to do the integral over $y$; only the lower limit is relevant:
$$I(q,k)=4{{\Gamma(3-{d\over2})}\over{(4\pi)^{d\over2}}}\int_0^1dx
x^{{d\over2}-1}(1-x)^{{d\over2}-2}{{(k^2)^{{d\over2}-1}}\over{(2-{d\over2})}}
\left({1\over{q^2+4x(1-x)k^2}}+O\left({1\over q^4}\right)\right).\eqno(A.7)$$
The remaining integral over $x$, and the subsequent integral over $q$ are now
straightforward, the final contribution to (2.29) being
$$\eqalign{J(k)&=-{{Z_\sigma g^2}\over N_f}\int_q{A_d\over{(q^2)^{{d\over2}-1}
+\mu^{d-2}}}I(q,k)\cr
&=-8{{Z_\sigma g^2}\over N_f}{{\Gamma(2-{d\over2})B({d\over2},{d\over2}-1)A_d}
\over{(4\pi)^d\Gamma({d\over2})}}(k^2)^{{d\over2}-1}\ln{\Lambda\over{\{\mu\vert
k\}}}.\cr}\eqno(A.8)$$
\vfill\eject
\noindent{\bf Acknowledgements}

SJH is partly supported by an SERC Advanced Fellowship. JBK is partially
supported by the National Science Foundation, NSF-PHY97-00148. The computer
simulations reported here used the facilities of the Materials Research
Laboratory of the University of Illinois, NSF-MR189-20358538, as well as the
supercomputer centers at the University of Pittsburgh, the National Energy
Research Supercomputer Center, and Florida State University.
We have benefitted from discussing various aspects of this work and related
topics with Chris Allton, Nick Dorey,
Gil Gat, John Gracey,
Peter Hasenfratz, Tony Kennedy, Ely Klepfish, Alex Kovner, Ian Lawrie,
Martin L\"uscher, Brian Pendleton, Ray Renken, Baruch Rosenstein and Alan
Scheinine.
\vfill\eject
\centerline{\bf References}

\noindent
[1] Y. NAMBU AND G. JONA-lASINIO, {\sl Phys. Rev.\/} {\bf122} (1961), 345.

\noindent
[2] D. LURI\'E AND A.J. MACFARLANE, {\sl Phys. Rev.\/} {\bf136} (1964), B816.

\noindent
[3] K. TAMVAKIS AND G.S. GURALNIK, {\sl Nucl. Phys.\/} {\bf B146} (1978), 224.

\noindent
[4] T. EGUCHI, {\sl Phys. Rev.\/} {\bf D17} (1978), 611;\hfill\break
K.-I. SHIZUYA, {\sl Phys. Rev.\/} {\bf D21} (1980), 2327.

\noindent
[5] D.J. GROSS AND A. NEVEU, {\sl Phys. Rev.\/} {\bf D10} (1974), 3235.

\noindent
[6] K.G. WILSON, {\sl Phys. Rev.\/} {\bf D7} (1973), 2911.

\noindent
[7] G. PARISI, {\sl Nucl. Phys.\/} {\bf B100} (1975),368;\hfill\break
D.J. GROSS, {\sl in\/} ``Methods in Field Theory '' (Les Houches XXVIII)
(R. Balian
and J. Zinn-Justin, Eds.), North Holland, Amsterdam, 1976.

\noindent
[8] S.J. HANDS, A. KOCI\'C AND J.B. KOGUT, {\sl Phys. Lett.\/} {\bf B273}
(1991), 111.

\noindent
[9] B. ROSENSTEIN, B.J. WARR AND S.H. PARK, {\sl Phys. Rev. Lett.\/} {\bf62}
(1989), 1433; {\sl Phys. Rep.\/} {\bf205} (1991), 205.

\noindent
[10] Y. KIKUKAWA AND K. YAMAWAKI, {\sl Phys. Lett.\/} {\bf B234} (1990), 497.

\noindent
[11] G. GAT, A. KOVNER, B. ROSENSTEIN AND B.J. WARR, {\sl Phys. Lett.\/}
{\bf B240} (1990), 158.

\noindent
[12] C. DE CALAN, P.A. FARIA DA VEIGA, J. MAGNEN AND R. S\'EN\'EOR, {\sl Phys.
Rev. Lett.\/} {\bf66} (1991), 3233.

\noindent
[13] S. WEINBERG, {\sl Phys. Rev.\/} {\bf130} (1963), 776.

\noindent
[14] L.P. KADANOFF, {\sl Physics\/} {\bf2} (1966), 263.

\noindent
[15] G.A. BAKER JR., {\sl Phys. Rev. Lett.\/} {\bf20} (1968), 990.

\noindent
[16] A. KOCI\'C, {\sl Phys. Lett.\/} {\bf B281} (1992), 309.

\noindent
[17] B. ROSENSTEIN, B.J. WARR AND S.H. PARK, {\sl Phys. Rev.\/} {\bf D39}
(1989), 3088.

\noindent
[18] D.J. AMIT, ``Field Theory, the Renormalization Group, and Critical
Phenomena'', World Scientific, Singapore, 1984;\hfill\break
V.A. MIRANSKY, {\sl in \/} ``Proceedings of the 1990 International Workshop on
Strong Coupling Gauge Theories and Beyond'' (Nagoya, Japan, 1990) (T. Muta
and K. Yamawaki, Eds.), World Scientific, Singapore, 1991.

\noindent
[19] V.G. KOURES AND K.T. MAHANTHAPPA, {\sl Phys. Rev.\/} {\bf D45} (1992),
580.

\noindent
[20] S.-K. MA, ``Modern Theory of Critical Phenomena'', W.A. Benjamin Inc.,
Reading Mass., 1976.

\noindent
[21] S. HIKAMI AND T. MUTA, {\sl Prog. Theo. Phys.\/} {\bf57} (1977), 785.

\noindent
[22] G. GAT, A. KOVNER AND B. ROSENSTEIN, Vancouver preprint UBCTP-91-021,
1991.

\noindent
[23] J.A. GRACEY, {\sl Int. J. Mod. Phys.\/} {\bf A6} (1991), 395, 2755(E).

\noindent
[24] J. ZINN-JUSTIN, {\sl Nucl. Phys.\/} {\bf B367} (1991), 105;\hfill\break
M. CAMPOSTRINI AND P. ROSSI, {\sl Int. J. Mod. Phys.\/} {\bf A7} (1992), 3265.

\noindent
[25] M. GELL-MANN AND M. L\'EVY, {\sl Nuov. Cim.\/} {\bf16} (1960), 705.

\noindent
[26] S.-K. MA, {\sl Phys. Rev.\/} {\bf A7} (1973), 2172;\hfill\break
K. SYMANZIK, ``$1/N$ Expansion in $P(\phi^2)_{4-\epsilon}$ Theory
I. Massless Theory $0<\epsilon<2$'', DESY preprint 77/05, 1977.

\noindent
[27] Y. COHEN, S. ELITZUR AND E. RABINOVICI, {\sl Nucl. Phys.\/} {\bf B220}
(1983), 102.

\noindent
[28] C.J. BURDEN AND A.N. BURKITT, {\sl Europhys. Lett.\/} {\bf3} (1987), 545.

\noindent
[29] TH. JOLIC\OE UR, {\sl Phys. Lett.\/} {\bf171B} (1986), 431;\hfill\break
TH. JOLIC\OE UR, A. MOREL and B. PETERSSON, {\sl Nucl. Phys.\/} {\bf B274}
(1986), 225.

\noindent
[30] S. DUANE, A.D. KENNEDY, B.J. PENDLETON AND D. ROWETH, {\sl Phys. Lett.\/}
{\bf B195} (1987), 216.

\noindent
[31] L. B\'ELANGER, R. LACAZE, A. MOREL, N. ATTIG, B. PETERSSON AND
M. WOLFF, {\sl Nucl. Phys.\/} {\bf B340} (1990), 245.

\noindent
[32] S.J. HANDS, A. KOCI\'C AND J.B. KOGUT, Illinois preprint ILL-TH-92-\#13,
1992.

\noindent
[33] S. GOTTLIEB, W. LIU, D. TOUSSAINT, R.L. RENKEN AND R.L. SUGAR, {\sl Phys.
Rev.\/} {\bf D38} (1988), 2245.

\noindent
[34] M. ABRAMOWITZ AND I.A. STEGUN, ``Handbook of Mathematical Functions'',
chap. 5, Dover, New York, 1972.

\noindent
[35] A.V. BALATSKY, {\sl Phys. Rev. Lett.\/} {\bf64} (1990), 2078.

\noindent
[36] S.J. HANDS, A. KOCI\'C, J.B. KOGUT, R.L. RENKEN, D.K. SINCLAIR AND
K.C. WANG, {\sl Phys. Lett.\/} {\bf B261} (1991), 294;
Illinois preprint ILL-TH-92-\#16, 1992;\hfill\break
A. KOCI\'C, J.B. KOGUT AND K.C. WANG, Illinois preprint ILL-TH-92-\#18, 1992.
\vfill\eject
\centerline{\bf Table I}
\vskip 0.5 truecm
\noindent
Table of critical exponents and critical couplings
$1/G_c^2=1/g_c^2\times(4\pi)^{d\over2}\Gamma({d\over2})(d-2)/8\Lambda^
{d-2}$ for the three variants of the Gross-Neveu model considered in
section 2. The constant $C_d$ is defined in Eq.(2.13).
\vskip 1 truecm
$$\vbox{\settabs\+${\displaystyle{1\over G_c^2}}$&\qquad\qquad&
\qquad${\displaystyle
      {1\over(d-2)}\left[1+{(d-1)\over d}{C_d\over N_f}\right]}$&
\qquad${\displaystyle
      {1\over(d-2)}\left[1+{{2(d-1)}\over d}{C_d\over N_f}\right]}$&
\qquad${\displaystyle
      {1\over(d-2)}\left[1+{{2(d-1)}\over d}{C_d\over N_f}\right]}$&\cr
\+&&\hfill ${\rm Z}_2$\hfill&\hfill${\rm U(1)}_L\otimes{\rm U(1)}_R$\hfill&
\hfill${\rm SU(2)}_L\otimes{\rm SU(2)}_R$\hfill&\cr
\bigskip
\bigskip
\bigskip
\+\hfill$\beta$\hfill&&\hfill${\displaystyle{1\over(d-2)}}$\hfill&
\hfill${\displaystyle
       {1\over(d-2)}\left[1+{C_d\over N_f}\right]}$\hfill&
\hfill${\displaystyle
       {1\over(d-2)}\left[1+{3\over2}{C_d\over N_f}\right]}$\hfill&\cr
\bigskip
\bigskip
\+\hfill$\delta$\hfill&&
\hfill${\displaystyle
       (d-1)\left[1+{C_d\over N_f}\right]}$\hfill&
\hfill${\displaystyle
       (d-1)\left[1+{(d-2)\over(d-1)}{C_d\over N_f}\right]}$\hfill&
\hfill${\displaystyle
       (d-1)\left[1+{(d-4)\over{2(d-1)}}{C_d\over N_f}\right]}$\hfill&\cr
\bigskip
\bigskip
\+\hfill$\gamma$\hfill&&
\hfill${\displaystyle
       1+{(d-1)\over(d-2)}{C_d\over N_f}}$\hfill&
\hfill${\displaystyle
       1+2{C_d\over N_f}}$\hfill&
\hfill${\displaystyle
       1+{(2d-5)\over(d-2)}{C_d\over N_f}}$\hfill&\cr
\bigskip
\bigskip
\+\hfill$\nu$\hfill&&
\qquad ${\displaystyle
       {1\over(d-2)}\left[1+{(d-1)\over d}{C_d\over N_f}\right]}$&
\qquad ${\displaystyle
      {1\over(d-2)}\left[1+{{2(d-1)}\over d}{C_d\over N_f}\right]}$&
\qquad ${\displaystyle
      {1\over(d-2)}\left[1+{{2(d-1)}\over d}{C_d\over N_f}\right]}$&\cr
\bigskip
\bigskip
\+\hfill$\eta$\hfill&&
\hfill${\displaystyle
       4-d-{{2(d-1)}\over d}{C_d\over N_f}}$\hfill&
\hfill${\displaystyle
       4-d-{{2(d-2)}\over d}{C_d\over N_f}}$\hfill&
\hfill${\displaystyle
       4-d+{(4-d)\over d}{C_d\over N_f}}$\hfill&\cr
\bigskip
\bigskip
\bigskip
\bigskip
\+${\displaystyle
   {1\over G_c^2}}$&&
\hfill${\displaystyle
       1-{(d-1)\over2N_f}}$\hfill&
\hfill${\displaystyle
        1-{(d-2)\over N_f}}$\hfill&
\hfill${\displaystyle
       2\left[1-{(2d-5)\over2N_f}\right]}$\hfill&\cr}$$
\vfill\eject
\centerline{\bf Table II}
\vskip 0.5 truecm
\noindent
Vacuum expectation value $\Sigma_0(N_f=\infty)$ as predicted by the
leading order lattice gap equation, together with measured deviations
$N_f(\Sigma_0(\infty)-\Sigma_0(N_f))$ vs. coupling $1/g^2$, for
$N_f=6,12,24$.
\vskip 1 truecm
$$\vbox{\settabs\+\quad  .5\quad &\quad .8351\quad &\quad
$24(\Sigma_0(\infty)-\Sigma_0(24))$\quad &\quad
$12(\Sigma_0(\infty)-\Sigma_0(12))$\quad &\quad
$ 6(\Sigma_0(\infty)-\Sigma_0( 6))$\quad &\cr
\+\hfill${1/g^2}$\hfill&\hfill$\Sigma_0(\infty)$\hfill&\quad
$24(\Sigma_0(\infty)-\Sigma_0(24))$\quad &\quad
$12(\Sigma_0(\infty)-\Sigma_0(12))$\quad &\quad
$ 6(\Sigma_0(\infty)-\Sigma_0( 6))$\quad &\cr\bigskip
\+\quad  .5\quad &\quad .8351\quad &\hfill.199(14)\hfill&\hfill.227(12)\hfill&
\hfill.267(10)\hfill&\cr\smallskip
\+\hfill .6\hfill&\hfill.6401\hfill&\hfill.310(19)\hfill&\hfill.319(12)\hfill&
\hfill.382(11)\hfill&\cr\smallskip
\+\hfill .7\hfill&\hfill.4711\hfill&\hfill.408(22)\hfill&\hfill.402(13)\hfill&
\hfill.517(15)\hfill&\cr\smallskip
\+\hfill .8\hfill&\hfill.3165\hfill&\hfill.542(34)\hfill&\hfill.550(24)\hfill&
\hfill.634(30)\hfill&\cr}$$
\vfill\eject
\centerline{\bf Table III}
\vskip 0.5 truecm
\noindent
Vacuum expectation value $\Sigma_0$ and its susceptibility
$\chi$ vs. coupling $1/g^2$ on 	a
$8^3$ lattice.  The number of trajectories of the hybrid Monte Carlo
algorithm at each value of $1/g^2$ are listed in column 4.
\vskip 1 truecm
$$\vbox{\settabs\+\qquad1.025\qquad&\qquad.4397(12)\qquad&\qquad1.485(35)\qquad
&\qquad Trajectories\qquad&\cr
\+\hfill${1/g^2}$\hfill&\hfill$\Sigma_0$\hfill&\hfill$\chi$\hfill&
\qquad Trajectories\qquad&\cr
\bigskip
\+\hfill.70\hfill&\qquad.4397(12)\qquad&\hfill.265(8)\hfill&\hfill5,000\hfill&
\cr\smallskip
\+\hfill.725\hfill&\hfill.3974(14)\hfill&\hfill.308(12)\hfill&\hfill5,000\hfill
&\cr
\smallskip
\+\hfill.75\hfill&\hfill.3544(16)\hfill&\hfill.400(14)\hfill&\hfill5,000\hfill
&\cr
\smallskip
\+\hfill.775\hfill&\hfill.3114(20)\hfill&\hfill.534(25)\hfill&\hfill5,000\hfill
&\cr
\smallskip
\+\hfill.80\hfill&\hfill.2693(20)\hfill&\hfill.721(31)\hfill&\hfill7,500\hfill
&\cr
\smallskip
\+\hfill.825\hfill&\hfill.2136(21)\hfill&\qquad1.485(35)\qquad&\hfill7,500
\hfill&\cr\smallskip
\+\hfill.85\hfill&\hfill.1579(19)\hfill&\hfill2.07(14)\hfill&\hfill10,000\hfill
&\cr
\smallskip
\+\hfill.875\hfill&\hfill.1152(19)\hfill&\hfill1.75(18)\hfill&\hfill10,000\hfill
&\cr
\smallskip
\+\hfill.90\hfill&\hfill--\hfill&\hfill1.38(18)\hfill&\hfill10,000\hfill&\cr
\smallskip
\+\hfill.925\hfill&\hfill--\hfill&\hfill.916(44)\hfill&\hfill10,000\hfill&\cr
\smallskip
\+\hfill.95\hfill&\hfill--\hfill&\hfill.633(41)\hfill&\hfill7,500\hfill&\cr
\smallskip
\+\hfill.975\hfill&\hfill--\hfill&\hfill.489(32)\hfill&\hfill7,500\hfill&\cr
\smallskip
\+\hfill1.00\hfill&\hfill--\hfill&\hfill.363(21)\hfill&\hfill5,000\hfill&\cr
\smallskip
\+\qquad1.025\qquad&\hfill--\hfill&\hfill.270(9)\hfill&\hfill5,000\hfill&\cr
\smallskip
\+\hfill1.05\hfill&\hfill--\hfill&\hfill.215(7)\hfill&\hfill5,000\hfill&\cr}$$
\vfill\eject
\centerline{\bf Table IV}
\vskip 0.5 truecm
\noindent
Vacuum expectation value $\Sigma_0$ and its susceptibility
$\chi$ vs. coupling $1/g^2$ on 	a
$12^3$ lattice.
\vskip 1 truecm
$$\vbox{\settabs\+\qquad1.025\qquad&\qquad.4319(10)\qquad&\qquad6.38(2.00)
\qquad&\qquad Trajectories\qquad&\cr
\+\hfill${1/g^2}$\hfill&\hfill$\Sigma_0$\hfill&\hfill$\chi$\hfill&
\qquad Trajectories\qquad&\cr
\bigskip
\+\hfill.70\hfill&\qquad.4319(10)\qquad&\hfill.511(11)\hfill&\hfill5,000\hfill&
\cr\smallskip
\+\hfill.725\hfill&\hfill.3905(12)\hfill&\hfill.621(13)\hfill&\hfill5,000\hfill
&\cr
\smallskip
\+\hfill.75\hfill&\hfill.3482(12)\hfill&\hfill.695(14)\hfill&\hfill5,000\hfill
&\cr
\smallskip
\+\hfill.775\hfill&\hfill.3091(13)\hfill&\hfill.876(25)\hfill&\hfill5,000\hfill
&\cr
\smallskip
\+\hfill.80\hfill&\hfill.2655(15)\hfill&\hfill.974(33)\hfill&\hfill5,000\hfill
&\cr
\smallskip
\+\hfill.825\hfill&\hfill.2255(15)\hfill&\hfill1.06(7)\hfill&\hfill7,000\hfill
&\cr
\smallskip
\+\hfill.85\hfill&\hfill.1786(17)\hfill&\hfill2.34(17)\hfill&\hfill7,000\hfill
&\cr
\smallskip
\+\hfill.875\hfill&\hfill.1262(21)\hfill&\hfill3.28(75)\hfill&\hfill10,000\hfill
&\cr
\smallskip
\+\hfill.90\hfill&\hfill.0853(25)\hfill&\hfill3.38(81)\hfill&\hfill10,000\hfill
&\cr
\smallskip
\+\hfill.925\hfill&\hfill--\hfill&\qquad6.38(2.00)\qquad&\hfill10,000\hfill&\cr
\smallskip
\+\hfill.95\hfill&\hfill--\hfill&\hfill3.75(1.00)\hfill&\hfill10,000\hfill&\cr
\smallskip
\+\hfill.975\hfill&\hfill--\hfill&\hfill2.18(22)\hfill&\hfill5,000\hfill&\cr
\smallskip
\+\hfill1.00\hfill&\hfill--\hfill&\hfill1.87(21)\hfill&\hfill5,000\hfill&\cr
\smallskip
\+\qquad1.025\qquad&\hfill--\hfill&\hfill1.53(16)\hfill&\hfill5,000\hfill&\cr
\smallskip
\+\hfill1.05\hfill&\hfill--\hfill&\hfill1.22(4)\hfill&\hfill5,000\hfill&\cr}$$
\vfill\eject
\centerline{\bf Table V}
\vskip 0.5 truecm
\noindent
Vacuum expectation value $\Sigma_0$ and its susceptibility
$\chi$ vs. coupling $1/g^2$ on 	a
$16^3$ lattice.
\vskip 1 truecm
$$\vbox{\settabs\+\qquad1.025\qquad&\qquad.4323(15)\qquad&\qquad4.12(1.12)
\qquad&\qquad Trajectories\qquad&\cr
\+\hfill${1/g^2}$\hfill&\hfill$\Sigma_0$\hfill&\hfill$\chi$\hfill&
\qquad Trajectories\qquad&\cr
\bigskip
\+\hfill.70\hfill&\qquad.4323(15)\qquad&\hfill.546(24)\hfill&\hfill1,000\hfill&
\cr\smallskip
\+\hfill.725\hfill&\hfill.3888(15)\hfill&\hfill.556(25)\hfill&\hfill1,000\hfill
&\cr
\smallskip
\+\hfill.75\hfill&\hfill.3462(17)\hfill&\hfill.741(27)\hfill&\hfill1,000\hfill
&\cr
\smallskip
\+\hfill.775\hfill&\hfill.3065(18)\hfill&\hfill.826(45)\hfill&\hfill1,000\hfill
&\cr
\smallskip
\+\hfill.80\hfill&\hfill.2625(21)\hfill&\hfill.988(60)\hfill&\hfill1,000\hfill
&\cr
\smallskip
\+\hfill.825\hfill&\hfill.2219(22)\hfill&\hfill1.19(8)\hfill&\hfill1,000\hfill
&\cr
\smallskip
\+\hfill.85\hfill&\hfill.1747(23)\hfill&\hfill1.85(30)\hfill&\hfill2,000\hfill
&\cr
\smallskip
\+\hfill.875\hfill&\hfill.1306(25)\hfill&\hfill3.05(35)\hfill&\hfill2,000\hfill
&\cr
\smallskip
\+\hfill.90\hfill&\hfill.0885(35)\hfill&\qquad4.12(1.12)\qquad&\hfill2,000
\hfill&\cr\smallskip
\+\hfill.925\hfill&\hfill--\hfill&\hfill10.6(4.0)\hfill&\hfill2,000\hfill&\cr
\smallskip
\+\hfill.95\hfill&\hfill--\hfill&\hfill4.97(1.60)\hfill&\hfill2,000\hfill&\cr
\smallskip
\+\hfill.975\hfill&\hfill--\hfill&\hfill2.94(50)\hfill&\hfill1,000\hfill&\cr
\smallskip
\+\hfill1.00\hfill&\hfill--\hfill&\hfill3.22(50)\hfill&\hfill1,000\hfill&\cr
\smallskip
\+\qquad1.025\qquad&\hfill--\hfill&\hfill1.22(14)\hfill&\hfill1,000\hfill&\cr
\smallskip
\+\hfill1.05\hfill&\hfill--\hfill&\hfill.97(8)\hfill&\hfill1,000\hfill&\cr}$$
\vfill\eject
\centerline{\bf Table VI}
\vskip 0.5 truecm
\noindent
Vacuum expectation value $\Sigma_0$ and its susceptibility
$\chi$ vs. coupling $1/g^2$ on 	a
$20^3$ lattice.
\vskip 1 truecm
$$\vbox{\settabs\+\qquad1.025\qquad&\qquad.0873(15)\qquad&\qquad4.07(1.15)
\qquad&\qquad Trajectories\qquad&\cr
\+\hfill${1/g^2}$\hfill&\hfill$\Sigma_0$\hfill&\hfill$\chi$\hfill&
\qquad Trajectories\qquad&\cr
\bigskip
\+\hfill.70\hfill&\hfill.4317(2)\hfill&\hfill.611(15)\hfill&\hfill5,000\hfill
&\cr
\smallskip
\+\hfill.725\hfill&\hfill.3883(2)\hfill&\hfill.633(15)\hfill&\hfill5,000\hfill
&\cr
\smallskip
\+\hfill.75\hfill&\hfill.3460(2)\hfill&\hfill.746(25)\hfill&\hfill5,000\hfill
&\cr
\smallskip
\+\hfill.775\hfill&\hfill.3045(3)\hfill&\hfill.837(51)\hfill&\hfill5,000\hfill
&\cr
\smallskip
\+\hfill.80\hfill&\hfill.2616(3)\hfill&\hfill.948(55)\hfill&\hfill7,500\hfill
&\cr
\smallskip
\+\hfill.825\hfill&\hfill.2189(3)\hfill&\hfill1.21(8)\hfill&\hfill7,500\hfill
&\cr
\smallskip
\+\hfill.85\hfill&\hfill.1800(4)\hfill&\hfill1.40(9)\hfill&\hfill7,500\hfill
&\cr
\smallskip
\+\hfill.875\hfill&\hfill.1376(5)\hfill&\hfill2.24(15)\hfill&\hfill10,000\hfill
&\cr
\smallskip
\+\hfill.90\hfill&\qquad.0873(15)\qquad&\qquad4.07(1.15)\qquad&\hfill10,000
\hfill&\cr\smallskip
\+\hfill.925\hfill&\hfill.0406(20)\hfill&\hfill3.86(1.16)\hfill&\hfill10,000
\hfill&\cr
\smallskip
\+\hfill.95\hfill&\hfill--\hfill&\hfill7.49(2.50)\hfill&\hfill7,500\hfill&\cr
\smallskip
\+\hfill.975\hfill&\hfill--\hfill&\hfill4.19(1.20)\hfill&\hfill7,500\hfill&\cr
\smallskip
\+\hfill1.00\hfill&\hfill--\hfill&\hfill2.28(32)\hfill&\hfill7,500\hfill&\cr
\smallskip
\+\qquad1.025\qquad&\hfill--\hfill&\hfill1.66(15)\hfill&\hfill5,000\hfill&\cr
\smallskip
\+\hfill1.05\hfill&\hfill--\hfill&\hfill1.35(10)\hfill&\hfill5,000\hfill&\cr
\smallskip
\+\hfill1.075\hfill&\hfill--\hfill&\hfill1.05(7)\hfill&\hfill5,000\hfill&\cr
\smallskip
\+\hfill1.10\hfill&\hfill--\hfill&\hfill.770(2)\hfill&\hfill5,000\hfill&\cr}$$
\vfill\eject
\centerline{\bf Table VII}
\vskip 0.5 truecm
\noindent
Estimates of the bulk critical point on $L^3$ lattices with $L=8$, 12, 16,
20.
\vskip 1 truecm
$$\vbox{\settabs\+\qquad20\qquad&\qquad\qquad&\qquad.950(20)\qquad&\cr
\+\hfill$L$\hfill&&\hfill${1/g^2_c(L)}$\hfill&\cr
\bigskip
\+\hfill8\hfill&&\hfill.867(8)\hfill&\cr
\smallskip
\+\hfill12\hfill&&\hfill.925(20)\hfill&\cr
\smallskip
\+\hfill16\hfill&&\hfill.930(20)\hfill&\cr
\smallskip
\+\qquad20\qquad&&\qquad.950(20)\qquad&\cr}$$
\vfill
\centerline{\bf Table VIII}
\vskip 0.5 truecm
\noindent
Estimates of the finite temperature critical points on $N_\tau\times N^2$
lattices for $N_\tau=2$, 4, 6, 8, 10 and 12.
\vskip 1 truecm
$$\vbox{\settabs\+\qquad12\qquad&\qquad\qquad&\qquad.880(5)\qquad&\cr
\+\hfill$N_\tau$\hfill&&\hfill${1/g^2_{\beta c}}$\hfill&\cr
\bigskip
\+\hfill2\hfill&&\hfill.47(1)\hfill&\cr
\smallskip
\+\hfill4\hfill&&\hfill.695(5)\hfill&\cr
\smallskip
\+\hfill6\hfill&&\hfill.785(5)\hfill&\cr
\smallskip
\+\hfill8\hfill&&\hfill.833(5)\hfill&\cr
\smallskip
\+\hfill10\hfill&&\hfill.865(5)\hfill&\cr
\smallskip
\+\qquad12\qquad&&\qquad.880(20)\qquad&\cr}$$
\vfill\eject
\centerline{\bf Table IX}
\vskip 0.5 truecm
\noindent
Zero temperature vacuum expectation values of $\Sigma_0$ (from Fig. 9) at the
couplings of the finite temperature critical points recorded in Table VII,
and $T_c$ estimates for $N_\tau$ ranging from 2 through 12 in units of
$\Sigma_0$.
\vskip 1 truecm
$$\vbox{\settabs\+\qquad12\qquad&\qquad.125(10)\qquad&\qquad.67(5)\qquad&\cr
\+\hfill$N_\tau$\hfill&\hfill$\Sigma_0$\hfill&\hfill$T_c/\Sigma_0$\hfill&\cr
\bigskip
\+\hfill2\hfill&\hfill--\hfill&\hfill--\hfill&\cr
\smallskip
\+\hfill4\hfill&\hfill.44(1)\hfill&\hfill.57(1)\hfill&\cr
\smallskip
\+\hfill6\hfill&\hfill.29(1)\hfill&\hfill.57(2)\hfill&\cr
\smallskip
\+\hfill8\hfill&\hfill.21(1)\hfill&\hfill.60(3)\hfill&\cr
\smallskip
\+\hfill10\hfill&\hfill.157(10)\hfill&\hfill.64(4)\hfill&\cr
\smallskip
\+\qquad12\qquad&\qquad.125(10)\qquad&\qquad.67(5)\qquad&\cr}$$
\vfill\eject
\centerline{\bf Table X}
\vskip 0.5 truecm
\noindent
Vacuum expectation value $\Sigma_0$ vs. fermion bare mass $m$ at coupling
$1/g^2=1.00$ on $8^3$, $16^3$ and $24^3$ lattices. The number of trajectories
used for each measurement are given in the last column.
\vskip 1 truecm
$$\vbox{\settabs\+\qquad.01625\qquad&\qquad.0496(12)\qquad&\qquad.0614(10)
\qquad&\qquad.0594(10)\qquad&\qquad Trajectories\qquad&\cr
\+\hfill$m$\hfill&\hfill$\Sigma_0(8^3)$\hfill&\hfill$\Sigma_0(16^3)$\hfill&
\hfill$\Sigma_0(24^3)$\hfill&\qquad Trajectories\qquad&\cr
\bigskip
\+\hfill.005\hfill&\qquad.0496(12)\qquad&\qquad.0614(10)\qquad&\qquad.0594(10)
\qquad&\hfill10,000\hfill&\cr\smallskip
\+\hfill.008\hfill&\hfill.0703(13)\hfill&\hfill.0783(9)\hfill&\hfill.0814(10)
\hfill&\hfill10,000\hfill&\cr\smallskip
\+\hfill.0125\hfill&\hfill.0931(5)\hfill&\hfill.106(1)\hfill&\hfill.106(1)\hfill
&\hfill10,000\hfill&\cr\smallskip
\+\qquad.01625\qquad&\hfill.118(2)\hfill&\hfill.122(1)\hfill&\hfill.123(1)\hfill
&\hfill10,000\hfill&\cr\smallskip
\+\hfill.020\hfill&\hfill.138(1)\hfill&\hfill.138(1)\hfill&\hfill.137(1)\hfill&
\hfill10,000\hfill&\cr\smallskip
\+\hfill.025\hfill&\hfill.155(1)\hfill&\hfill.154(1)\hfill&\hfill.153(1)\hfill&
\hfill10,000\hfill&\cr\smallskip
\+\hfill.0375\hfill&\hfill.190(1)\hfill&\hfill.187(1)\hfill&\hfill.187(1)\hfill
&\hfill10,000\hfill&\cr\smallskip
\+\hfill.050\hfill&\hfill.218(1)\hfill&\hfill.213(1)\hfill&\hfill.213(1)\hfill&
\hfill5,000\hfill&\cr\smallskip
\+\hfill.0625\hfill&\hfill.239(1)\hfill&\hfill.234(1)\hfill&\hfill.235(1)\hfill&
\hfill5,000\hfill&\cr\smallskip
\+\hfill.075\hfill&\hfill.257(1)\hfill&\hfill.253(1)\hfill&\hfill.252(1)\hfill&
\hfill5,000\hfill&\cr\smallskip
\+\hfill.10\hfill&\hfill.287(1)\hfill&\hfill.282(1)\hfill&\hfill.282(1)\hfill&
\hfill5,000\hfill&\cr\smallskip
\+\hfill.15\hfill&\hfill.327(1)\hfill&\hfill--\hfill&\hfill--\hfill&\hfill5,000
\hfill&\cr\smallskip
\+\hfill.20\hfill&\hfill.355(1)\hfill&\hfill--\hfill&\hfill--\hfill&\hfill5,000
\hfill&\cr\smallskip
\+\hfill.25\hfill&\hfill.377(1)\hfill&\hfill--\hfill&\hfill--\hfill&\hfill5,000
\hfill&\cr}$$
\vfill\eject
\centerline{\bf Table XI}
\vskip 0.5 truecm
\noindent
Vacuum expectation value $\Sigma_0$ vs. fermion bare mass $m$ at coupling
$1/g^2=0.975$ on $8^3$, $16^3$ and $24^3$ lattices.
\vskip 1 truecm
$$\vbox{\settabs\+\qquad.01625\qquad&\qquad.0647(10)\qquad&\qquad.0705(5)\qquad
&\qquad.0767(3)\qquad&\qquad Trajectories\qquad&\cr
\+\hfill$m$\hfill&\hfill$\Sigma_0(8^3)$\hfill&\hfill$\Sigma_0(16^3)$\hfill&
\hfill$\Sigma_0(24^3)$\hfill&\qquad Trajectories\qquad&\cr
\bigskip
\+\hfill.005\hfill&\qquad.0647(10)\qquad&\qquad.0705(5)\qquad&\qquad.0767(3)
\qquad&\hfill10,000\hfill&\cr\smallskip
\+\hfill.008\hfill&\hfill.0943(8)\hfill&\hfill.0968(3)\hfill&\hfill.0980(3)
\hfill&\hfill10,000\hfill&\cr\smallskip
\+\hfill.0125\hfill&\hfill.118(2)\hfill&\hfill.123(1)\hfill&\hfill.122(1)\hfill
&\hfill10,000\hfill&\cr\smallskip
\+\qquad.01625\qquad&\hfill.134(1)\hfill&\hfill.135(1)\hfill&\hfill
.129(1)\hfill
&\hfill10,000\hfill&\cr\smallskip
\+\hfill.020\hfill&\hfill.150(1)\hfill&\hfill.152(1)\hfill&\hfill.152(1)\hfill&
\hfill10,000\hfill&\cr\smallskip
\+\hfill.025\hfill&\hfill.170(1)\hfill&\hfill.170(1)\hfill&\hfill.169(1)\hfill&
\hfill10,000\hfill&\cr\smallskip
\+\hfill.0375\hfill&\hfill.207(1)\hfill&\hfill.203(1)\hfill&\hfill.202(1)\hfill
&\hfill10,000\hfill&\cr\smallskip
\+\hfill.050\hfill&\hfill.233(1)\hfill&\hfill.229(1)\hfill&\hfill.228(1)\hfill&
\hfill5,000\hfill&\cr\smallskip
\+\hfill.0625\hfill&\hfill.254(1)\hfill&\hfill.250(1)\hfill&\hfill.249(1)\hfill&
\hfill5,000\hfill&\cr\smallskip
\+\hfill.075\hfill&\hfill.272(1)\hfill&\hfill.267(1)\hfill&\hfill.267(1)\hfill&
\hfill5,000\hfill&\cr\smallskip
\+\hfill.10\hfill&\hfill.300(1)\hfill&\hfill.297(1)\hfill&\hfill.297(1)\hfill&
\hfill5,000\hfill&\cr}$$
\vfill\eject
\centerline{\bf Table XII}
\vskip 0.5 truecm
\noindent
Vacuum expectation value $\Sigma$ vs. fermion bare mass $m$ on a $12\times36^2$
lattice at coupling $1/g^2=0.8775$.  The number of trajectories used for each
measurement are given in the last column.
\vskip 1 truecm
$$\vbox{\settabs\+\qquad.00125\qquad&\qquad.0903(10)\qquad&\qquad Trajectories
\qquad&\cr
\+\hfill$m$\hfill&\hfill$\Sigma$\hfill&\qquad Trajectories\qquad&\cr
\bigskip
\+\hfill.01\hfill&\hfill.1826(8)\hfill&\hfill1,000\hfill&\cr\smallskip
\+\hfill.005\hfill&\hfill.1471(4)\hfill&\hfill1,000\hfill&\cr\smallskip
\+\hfill.0025\hfill&\hfill.1151(2)\hfill&\hfill1,000\hfill&\cr\smallskip
\+\hfill.0018\hfill&\hfill.1084(2)\hfill&\hfill1,000\hfill&\cr\smallskip
\+\qquad.00125\qquad&\qquad.0903(10)\qquad&\hfill1,200\hfill&\cr\smallskip
\+\hfill.0009\hfill&\hfill.0805(10)\hfill&\hfill1,200\hfill&\cr}$$
\vfill\eject
\vfill\eject
\centerline{\bf Table XIII}

\noindent
Results of fits to the scalar correlator in the broken phase, using both
the branch cut form $P(x;M)$ and the usual simple pole form.
\vskip 1 truecm
$$\vbox{\settabs\+\quad.775\quad&\quad\quad&
\quad.388(58)\quad&\quad.092382(18)\quad
&\quad3.3\quad&\quad\quad&\quad.513(51)\quad&\quad.092391(15)\quad&
\quad5.1\quad&\cr
\+&&&\hfill branch cut\hfill&&&&\hfill simple pole\hfill&&\cr\bigskip
\+\hfill$1/g^2$\hfill&&\hfill$M$\hfill&\hfill$\Sigma_0^2$\hfill&\hfill$\chi^2$
\hfill&&\hfill$M$\hfill&\hfill$\Sigma_0^2$\hfill&\hfill$\chi^2$\hfill&\cr
\bigskip
\+\hfill.7\hfill&&\hfill.708(98)\hfill&\hfill.18606(1)\hfill&\hfill3.3\hfill&&
\hfill.812(94)\hfill&\hfill.18606(1)\hfill&\hfill2.9\hfill&\cr\smallskip
\+\hfill.725\hfill&&\hfill.750(95)\hfill&\hfill.15069(1)\hfill&\hfill4.2\hfill&&
\hfill.836(87)\hfill&\hfill.15069(1)\hfill&\hfill4.3\hfill&\cr\smallskip
\+\hfill.75\hfill&&\hfill.494(69)\hfill&\hfill.11958(1)\hfill&\hfill3.4\hfill&&
\hfill.607(64)\hfill&\hfill.11958(1)\hfill&\hfill3.7\hfill&\cr\smallskip
\+\quad.775\quad&&\quad.388(58)\quad&\quad.092382(18)\quad&
\quad3.3\quad&&\quad.513(51)\quad&\quad.092391(15)\quad&
\quad5.1\quad&\cr\smallskip
\+\hfill.8\hfill&&\hfill.433(58)\hfill&\hfill.068532(17)\hfill&\hfill6.2\hfill&&
\hfill.553(51)\hfill&\hfill.068539(15)\hfill&\hfill5.9\hfill&\cr\smallskip
\+\hfill.825\hfill&&\hfill.348(48)\hfill&\hfill.047656(23)\hfill&\hfill8.2\hfill
&&\hfill.478(40)\hfill&\hfill.047667(17)\hfill&\hfill6.7\hfill&\cr\smallskip
\+\hfill.85\hfill&&\hfill.320(57)\hfill&\hfill.032503(30)\hfill&\hfill2.1\hfill
&&\hfill.450(50)\hfill&\hfill.032519(22)\hfill&\hfill1.9\hfill&\cr\smallskip
\+\hfill.875\hfill&&\hfill.290(36)\hfill&\hfill.018956(25)\hfill&\hfill3.1\hfill
&&\hfill.433(27)\hfill&\hfill.018982(16)\hfill&\hfill5.1\hfill&\cr\smallskip
\+\hfill.9\hfill&&\hfill.281(42)\hfill&\hfill.007620(32)\hfill&\hfill4.7\hfill
&&\hfill.428(29)\hfill&\hfill.007653(19)\hfill&\hfill2.5\hfill&\cr\smallskip
\+\hfill.925\hfill&&\hfill.247(55)\hfill&\hfill.001906(46)\hfill&\hfill3.1\hfill
&&\hfill.387(39)\hfill&\hfill.001941(23)\hfill&\hfill2.4\hfill&\cr\smallskip
\+\hfill.95\hfill&&\hfill.354(57)\hfill&\hfill.000812(24)\hfill&\hfill2.0\hfill
&&\hfill.475(51)\hfill&\hfill.000821(19)\hfill&\hfill2.9\hfill&\cr\smallskip
\+\hfill.975\hfill&&\hfill.312(61)\hfill&\hfill.000383(29)\hfill&\hfill3.1\hfill
&&\hfill.459(51)\hfill&\hfill.000407(19)\hfill&\hfill2.5\hfill&\cr}$$
\vfill\eject
\centerline{\bf Table XIV}
\noindent
Results of fitting the scalar correlator in the symmetric phase to the
form $Q(x;\mu)$.
\vskip 1 truecm
$$\vbox{\settabs\+\qquad1.025\qquad&\qquad.156(13)\qquad&\qquad4.6\qquad&\cr
\+\hfill$1/g^2$\hfill&\hfill$\mu$\hfill&\hfill$\chi^2$\hfill&\cr
\bigskip
\+\hfill1.1\hfill&\hfill.437(55)\hfill&\hfill5.8\hfill&\cr\smallskip
\+\hfill1.075\hfill&\hfill.283(33)\hfill&\hfill3.1\hfill&\cr\smallskip
\+\hfill1.05\hfill&\hfill.159(14)\hfill&\hfill9.5\hfill&\cr\smallskip
\+\qquad1.025\qquad&\qquad.156(13)\qquad&\qquad4.6\qquad&\cr\smallskip
\+\hfill1.0\hfill&\hfill.146(12)\hfill&\hfill4.0\hfill&\cr\smallskip
\+\hfill0.975\hfill&\hfill.075(5)\hfill&\hfill4.7\hfill&\cr\smallskip
\+\hfill0.95\hfill&\hfill.041(2)\hfill&\hfill8.3\hfill&\cr\smallskip
\+\hfill0.925\hfill&\hfill.019(1)\hfill&\hfill4.8\hfill&\cr}$$
\vfill\eject
\centerline{\bf Figure Captions}

\noindent
{\bf Figure 1}: $O(1/N_f)$ corrections   to the fermion self-energy and
the fermion-scalar vertex.

\noindent{\bf Figure 2}: $O(1/N_f)$ correction to the gap equation.

\noindent{\bf Figure 3}: $O(1/N_f)$ corrections to the scalar propagator.

\noindent{\bf Figure 4}: Schematic diagram showing $1/N_f$ corrections to
the four-fermion scattering amplitude.

\noindent{\bf Figure 5}: Plot of $\Sigma_0$ vs. $1/g^2$ on a $12^3$ lattice,
for $N_f=24$ (squares), $N_f=12$ (circles), and $N_f=6$ (triangles). The solid
line is the leading order solution to the lattice gap equation (3.13).
Errors are smaller than the size of the symbols.

\noindent{\bf Figure 6}:
Plot of the data recorded in Table III.  $\chi^{-1}$, the reciprocal of the
susceptibility, is plotted multiplied by $10^{-1}$.

\noindent{\bf Figure 7}:
Plot of the data recorded in Table IV.  $\chi^{-1}$
is plotted multiplied by $10^{-1}$.

\noindent{\bf Figure 8}:
Plot of the data recorded in Table V.  $\chi^{-1}$
is plotted multiplied by $10^{-1}$.

\noindent{\bf Figure 9}:
Plot of the data recorded in Table VI.  $\chi^{-1}$
is plotted multiplied by $10^{-1}$.

\noindent{\bf Figure 10}:
Plot of $\ln\Sigma_0$ vs. $\ln(1/g^2_c-1/g^2)$ from Table VI,
$1/g^2_c=0.955$.

\noindent{\bf Figure 11}:
Plot of the bulk critical point's dependence on lattize size, $1/g^2_c(L)$
vs. $1/L$.

\noindent{\bf Figure 12}:
Histograms of $\Sigma$ measurements on a $10\times30^2$ lattice for couplings
$1/g^2=0.870$, 0.865 and 0.860.

\noindent{\bf Figure 13}:
Plot of the finite temperature critical point's dependence on the
temporal extent $N_\tau$ of the asymmetric $N_\tau\times N^2$ lattice.

\noindent{\bf Figure 14}:
Plot of $\ln(1/g^2_c-1/g^2_{\beta c}(N_\tau))$
vs. $\ln N_\tau$ for three estimates
(0.995, 0.976,
0.950) of the infinite volume critical point.

\noindent{\bf Figure 15}:
Plot of $\ln\Sigma_0$ vs. $\ln m$  for $8^3$, $16^3$ and $24^3$ lattices at
$1/g^2=1.00$, an estimate of the bulk critical point.

\noindent{\bf Figure 16}:
Plot of $\ln\Sigma_0$ vs. $\ln m$  for $8^3$, $16^3$ and $24^3$ lattices at
$1/g^2=0.975$, an estimate of the bulk critical point.

\noindent{\bf Figure 17}:
Plot of $\ln\Sigma$ vs. $\ln m$ on a $12\times36^2$ lattice at $1/g^2=0.8775$,
the finite temperature critical point for this lattice size.

\noindent{\bf Figure 18}
Plot of fits of the inverse correlation length vs. $1/g^2$ for a $20^3$
lattice.
The pluses show the scalar mass $M$ in the broken phase obtained using a fit to
a branch cut (7.4), the stars show $M$ values obtained using a simple pole fit,
and
the circles show the 2 fermion threshold $2\Sigma_0$ obtained using the branch
cut fit. In the symmetric phase the crosses show the scalar width $\mu$
fitted using the form (7.8).
\vfill\end